\documentclass[12pt]{article} 

\usepackage{physics}
\usepackage{graphicx}
\usepackage{amsmath}
\usepackage{amssymb}
\usepackage{hyperref}

\usepackage[paperwidth=8.5in,paperheight=11in,
  textwidth=6.25in,textheight=9.1in,
  centering]{geometry}

\usepackage{tikz}
\usepackage{cite}
\usepackage{amscd}
\usepackage{setspace}
\usepackage{bbm}
\usepackage{dsfont}
\usepackage{cancel}
\usepackage{relsize}
\usepackage[utf8]{inputenc}
\usepackage{xcolor}
\usepackage{tocloft}
\usepackage{mathtools}
\usepackage{newtxtext,newtxmath}
\definecolor{myblue}{RGB}{40, 80, 140}   
\definecolor{mygray}{RGB}{100, 100, 100}

\begin{document}
\begin{titlepage}

\begin{flushright}

\end{flushright}
\vskip 3cm
\begin{center}
{\Large \bf 
Dynamical Generation of the VY Superpotential in $\mathcal{N}=1$ SYM:
A Higher-Form Perspective}
\vskip 2.0cm

 Wei Gu \\

\bigskip
\begin{tabular}{cc}
Zhejiang Institute of Modern Physics, School of Physics,  Zhejiang University\\
Hangzhou, Zhejiang 310058, China\\

 \end{tabular}

\vskip 1cm

\textbf{Abstract}
\end{center}

We present a semiclassical account of the Veneziano--Yankielowicz (VY) superpotential in four-dimensional $\mathcal{N}=1$ super Yang--Mills theory. Motivated by two-dimensional gauged linear sigma models, where superpotentials arise from vortex dynamics, we reinterpret domain walls as fundamental objects associated with higher-form gauge fields. In this formulation, the vacuum structure is encoded in a compact three-form gauge field, whose four-form flux labels topological sectors. In the presence of charged matter with total charge $N$, these sectors exhibit a natural $\mathbb{Z}_N$ structure, leading to a decomposition into $N$ semiclassical contributions. These contributions arise from Euclidean point-like configurations in the higher-form sector, analogous to fractional instantons. We show that these configurations provide the relevant non-perturbative contributions to the effective superpotential. Integrating out the associated degrees of freedom reproduces the VY superpotential in the infrared. This gives a semiclassical origin of the VY superpotential in terms of higher-form gauge dynamics.

\medskip
\noindent

\bigskip
\vfill
\end{titlepage}

\setcounter{tocdepth}{2}
\tableofcontents

\section{Introduction}\label{Int}
The Veneziano--Yankielowicz (VY) superpotential\cite{Veneziano:1982ah} provides a fundamental description of the infrared dynamics of four-dimensional $\mathcal{N}=1$ supersymmetric Yang--Mills theory. 
It captures key non-perturbative features of the theory, including the gaugino condensate, the discrete set of $N$ vacua, and the associated domain walls \cite{Dvali:1996xe}. 
Traditionally, the VY superpotential is understood as an effective description determined by holomorphy, anomalies, and symmetry considerations \cite{Seiberg:1994bz}, rather than as arising directly from a microscopic semiclassical mechanism. Despite its success, the underlying microscopic mechanism remains unclear, and the strong-coupling dynamics is typically treated as a black box. 

To provide a dynamical explanation, we turn to the ultraviolet description of the theory.  At first sight, one might expect that nonperturbative effects from Yang--Mills instantons could account for the generation of a superpotential. However, in ordinary instanton backgrounds, this expectation is obstructed by the presence of the extra fermionic zero modes. 
In particular, an $SU(N)$ instanton carries $2N$ gaugino zero modes, which cannot be saturated to generate a superpotential term.
Various generalizations have been proposed to circumvent this obstruction, including matter-induced superpotentials such as the ADS superpotential~\cite{Affleck:1983mk}, compactification on $\mathbb{R}^3 \times S^1$~\cite{Seiberg:1996nz,Davies:1999uw}, and configurations involving fractional instantons or monopole-instantons~\cite{Davies:1999uw, Unsal:2008ch}. 
While these approaches provide valuable insights into the nonperturbative dynamics, either these approaches do not lead to a direct derivation of the VY superpotential, or they rely on modified settings where the low-energy dynamics, and hence the vacuum structure, differs from that of the original four-dimensional theory.

We approach this problem from a semiclassical perspective. While four-dimensional SYM itself does not admit a straightforward semiclassical derivation, two-dimensional gauged linear sigma models (GLSMs) provide a controlled setting in which similar superpotentials arise dynamically. This offers a useful starting point for understanding the structure of the VY superpotential. In GLSMs, the twisted effective superpotential can be obtained in two complementary ways: by integrating out massive matter fields in the large-$\Sigma$ regime \cite{Witten:1993yc}. or from vortex configurations (i.e. two-dimensional instantons), which generate a Hori–Vafa-type superpotential \cite{ Hori:2000kt, Witten:1993yc}. The latter description makes it natural to interpret the resulting superpotential in terms of localized semiclassical contributions.
In this formulation, it becomes straightforward to further integrate out the matter fields, leading to the effective twisted superpotential
\[
\sim \Sigma (\log \Sigma - 1)\,.
\]
This provides insight into the dynamical origin of the VY superpotential in four-dimensional $\mathcal{N}=1$ SYM. 

The key observation is the role of the fundamental matter fields $\phi_i$ in two-dimensional GLSMs, as originally studied in \cite{Witten:1978bc}. In addition to generating the twisted effective superpotential, these fields act as charged operators that create domain walls connecting distinct vacua. Gauge invariance requires attaching Wilson lines extending to infinity, leading to a natural description in terms of polar variables,
\[
\rho\, \exp\!\left(i \varphi(x_0) + i Q \int_{\gamma} A \right),
\]
where $\gamma$ is a path ending at $x_0$.

Another useful insight comes from the recent study in \cite{Gu:2025gtb}, which showed that several different GLSMs can share the same twisted effective superpotential. Thus, the vacuum structure and the associated BPS domain walls are the same. 
The distinction lies in whether the BPS domain walls correspond to dynamical excitations or are realized as non-dynamical defects, also referred to as external probes. In Section \ref{Les:GLSM}, we review these aspects, along with other relevant concepts of two-dimensional GLSMs. Readers familiar with the subject may safely skip this section.

As mentioned above, BPS domain walls also exist in four-dimensional $\mathcal{N}=1$ SYM and can be defined as in \cite{Dvali:1996xe}. More specifically, the nonperturbative dynamics in 4D $\mathcal{N}=1$ SYM generates a gluino condensate,
\[
\langle \lambda\lambda\rangle_k \sim \Lambda^3 \exp\!\left(\frac{2\pi i k}{N}\right),
\qquad k=0,1,\dots,N-1.
\]
As a result, the anomaly-free discrete $R$-symmetry $\mathbb{Z}_{2N}$ is spontaneously broken to $\mathbb{Z}_2$, leading to $N$ degenerate vacua. For a more detailed summary, see Section \ref{4d:SYM}. 
The BPS domain walls arise as nonperturbative configurations interpolating between distinct vacua. 
In four-dimensional pure $\mathcal{N}=1$ SYM, they are not associated with elementary fields, but instead emerge as solitonic objects. 
At the same time, as emphasized in \cite{Witten:1997ep}, they can be viewed as fundamental objects in the effective description, providing an alternative perspective on their role in the low-energy dynamics. In certain string/M-theory embeddings, the BPS domain walls of four-dimensional pure $\mathcal{N}=1$ SYM admit a brane realization: they arise from higher-dimensional branes wrapping internal supersymmetric cycles and therefore appear as $(2+1)$-dimensional walls in the noncompact spacetime \cite{Witten:1997ep, Acharya:2001dz, Bandos:2019qok}. This provides a geometric explanation for why the walls behave as extended nonperturbative objects. In the string/M-theory framework, Witten computed the value of the superpotential at a given vacuum\cite{Witten:1997ep}, from which it follows that the superpotential is of the VY-type.

The question is whether such dynamical generation can be understood within a field-theoretic framework. In conventional field theory, the fundamental excitations of elementary fields are point particles rather than extended objects. This conclusion is too naive. In gauge theories, one can define extended objects such as Wilson lines or loops. One special property of two-dimensional domain walls is that they behave as particles.  This is a crucial ingredient in understanding quantum integrable systems arising from GLSMs, as investigated in \cite{Gu:2022ugf}. 
In this sense, it explains the origin of elementary fields associated with domain walls in two-dimensional gauge theories. 
In contrast, domain walls in four dimensions are $(2+1)$-dimensional objects. 
This suggests that the corresponding fundamental degrees of freedom should be described by higher-form gauge fields. 

Before proceeding, let us note that the present construction admits a natural interpretation in terms of an axionic degree of freedom. 
In that language, domain walls arise as configurations interpolating between different branches associated with the periodicity of the axion. 
In the dual formulation, the phase variable is replaced by higher-form gauge fields, and the coupling to extended objects such as domain walls is naturally encoded by higher-dimensional Wilson-type operators. 
One is naturally led to consider operators of the form
\[
\rho \, \exp\!\left(i \int_{\Sigma = \partial M} B_2 + Q \int_M A_3 \right),
\]
where $\Sigma$ is the boundary of a three-dimensional manifold $M$, and $B_2$ and $A_3$ are two-form and three-form gauge fields, respectively.
They are treated as elementary fields in our field-theoretic framework. It is natural to expect that the radial mode $\rho$ contributes to local massive fluctuations around the wall profile, while the protected wall charge is controlled by the topological interpolation of the ``phase", i.e., the $B_2$ gauge fields. For a single $\phi$ field, $\rho$ is massive via the Higgs mechanism.
For multiple fields, a dynamical mass gap is required.
These fields can be coupled to 4D $\mathcal{N}=1$ SYM, resulting in a theory that we call \textit{4D $\mathcal{N}=1$ SYM extended by higher-form fields}. It turns out to be much more transparent to study the theory in the dual variables
\[
z=\varrho-ia \,,
\]
where $a$ is the axion field with many applications, such as in phenomenology \cite{Gabadadze:2002ff, Dvali:2005an}.  More details of this theory will be discussed in Section~\ref{4d:VYS}. In Section~\ref{4d:DDWD}, we will see the fact that these fields are massive does not imply that their dynamics are trivial.
Including such fields supports 4D Euclidean saddle configurations (in a fashion similar to the 2D vortex equations in GLSMs, as listed in Appendix~B).
As in the work of Hori and Vafa in 2D~\cite{Hori:2000kt}, one can generate a non-perturbative contribution to the superpotential, which can be written in terms of dual variables:
\[
\sim e^{-Z} \,.
\]

Finally, one can show that the VY superpotential in four-dimensional $\mathcal{N}=1$ SYM can be derived by integrating out the $Z$-fields in a holomorphic scheme, as discussed in Section~\ref{4d:VYS}.
This provides a field-theoretic explanation for the origin of the VY superpotential.

Let us close the introduction by briefly commenting on the symmetry interpretation of our construction. 
In four dimensions, the three-form gauge field provides a natural higher-form description of the branch structure associated with the infrared chiral sector. Domain walls interpolate between distinct vacua and therefore provide the natural extended objects in this formulation. This structure admits an interpretation in the language of higher-form symmetries \cite{Gaiotto:2014kfa}. 
For a related study, see \cite{Anber:2024gis} and references therein. In the dual description, the same structure is encoded by an axionic degree of freedom, whose Peccei--Quinn-type shift symmetry captures the infrared physics. As discussed in~\cite{Dvali:2005an}, the PQ symmetry can be described in terms of an intrinsic two-form gauge symmetry. In this formulation, when a field of charge $N$ acquires a vacuum expectation value, the corresponding gauge structure is Higgsed, leaving a residual discrete $\mathbb{Z}_N$ gauge symmetry. This leads to multiple degenerate vacua related by discrete shifts of the axion field, with domain walls interpolating between them across their worldvolume. In the presence of multiple fields, one linear combination of phases is gauged and removed via the Higgs mechanism. This perspective provides a unified way to interpret the multiple semiclassical contributions appearing in our framework.

\section{Lessons from Two-Dimensional Gauge Theories} \label{Les:GLSM}
In this section, we review some properties of $\mathcal{N}=(2,2)$ gauged linear sigma models (GLSMs)~\cite{Witten:1993yc}, which will be useful for understanding four-dimensional physics.

It was shown in~\cite{Gu:2025gtb} that several abelian GLSMs can share the same IR twisted effective superpotential:
\begin{equation}\label{2d:TESP}
    W_{\rm eff}=-N\Sigma\left(\log\left(\frac{\Sigma}{\Lambda_{2d}}\right)-1\right),
\end{equation}
where $\Sigma$ is the superfield strength and $\Lambda_{2d}$ is the dynamical scale. 
These target spaces can be formally described as gauged weighted projective spaces $\mathbb{CWP}(w_1,\ldots,w_n)$, where $\sum_i w_i = N$. 
By imposing the condition that their dynamical scales are equal\footnote{In our convention, the dynamical scale is given by 
\[
\Lambda^N = \Lambda_{\rm UV}^N e^{-t_0} \prod_i w_i^{-w_i},
\]
where $\Lambda_{\rm UV}$ is the UV cutoff and $t_0$ is the bare complexified FI parameter.}, then different target spaces share exactly the same ground states. 
However, the spectrum of dynamical BPS domain walls depends on the specific theory.  In particular, each charged field\footnote{A Wilson line must be attached to render it gauge invariant.}, $\Phi_i$, can excite an associated domain wall. Two special cases are worth noting: the charge-$N$ Abelian Higgs model, in which the effective theory decomposes into $N$ disconnected sectors with no dynamical domain walls between them; and the GLSM for $\mathbb{CP}^{N-1}$, where each field generates a fundamental domain wall connecting adjacent ground states.

There are two ways to obtain equation~(\ref{2d:TESP}) in GLSMs.  The first is to integrate out the matter fields directly in the large-$\Sigma$ region, where they are heavy~\cite{Witten:1993yc}. 
The second is to follow the prescription proposed by Hori and Vafa~\cite{Hori:2000kt}: quantum corrections from vortices in GLSMs generate a non-perturbative twisted effective superpotential written in terms of dual field variables; integrating out these dual variables then reproduces equation~(\ref{2d:TESP}). The second approach appears more complicated. However, its advantage is that the entire procedure can be carried out within a holomorphic scheme. We now provide further details. 
\vspace{1em}

\subsection{GLSMs in Dual Variables}\label{GL:Mi}

The bosonic charged matter fields in GLSMs are complex scalars $\phi_i$. Their Lagrangian can be expressed in terms of the polar variables $(\rho_i, \varphi_i)$, defined by $\phi_i = \rho_i e^{i\varphi_i}$, as follows:
\begin{equation}\label{2d:BL} \sum_i-(\partial_{\mu}\rho_i)^2 -\rho_i^2(\partial_{\mu}\varphi_i+w_iA_{\mu})^2,
\end{equation}
where $w_i$ is the charge of $\phi_i$. 
In two dimensions, one can perform an Abelian duality transformation on the variable $\varphi_i$~\cite{Deligne:1999qp,Hori:2000kt}, introducing a dual dynamical theta-angle variable $\vartheta_i$, with the Lagrangian given by
\begin{equation}\label{2d:DL} \sum_i-(\partial_{\mu}\rho_i)^2 -\frac{1}{4\rho_i^2}(\partial_{\mu}\vartheta_i)^2-\frac{1}{2}w_i\vartheta_i\epsilon^{\mu\nu}F_{\mu\nu}.
\end{equation}
The supersymmetric completion of the last term is $w_i Y_i \Sigma$, where $Y_i$ is a dual twisted chiral multiplet. 
Its lowest component is $y_i = \varrho_i - i \vartheta_i$, which is related to the original field variables; for example, $\varrho_i = \rho_i^2$. 
The classical twisted superpotential is then
\begin{equation}\label{2d:CPS}
\left(\sum_i w_i Y_i - t\right)\Sigma.
\end{equation}
There are perturbative corrections to the parameter $t$ and the variables $Y_i$:
\begin{equation}\nonumber
t(\mu) = t_0 + N \log\frac{\mu}{\Lambda_{\rm UV}}, 
\quad 
Y_i = Y_{i0} + w_i \log\frac{\mu}{\Lambda_{\rm UV}}.
\end{equation}
In addition to perturbative corrections, one also expects instantons to contribute to the superpotential for the $Y$ fields. 
In two dimensions, vortices play the role of instantons. 
For example, for $N=1$, one expects a term
\begin{equation}\nonumber 
e^{-Y}.
\end{equation}
Such terms are generated in the superpotential. 
However, a puzzle arises: the numbers of bosonic and fermionic zero modes of a degree-one vortex are both $2N$, and thus such configurations cannot contribute to the superpotential unless $N=1$. 
Hori and Vafa resolved this issue by observing that one can embed a $U(1)$ gauge theory into a $U(1)^n$ gauge theory by gauging the $U(1)^n$ global symmetry, and recover the original theory in the weak-coupling limit of the additional gauge interactions. 
This leads to the exact superpotential
\begin{equation}\label{2d:MS} 
W=\left(\sum^{n}_{i=1} w_i Y_i - t\right)\Sigma+\sum^{n}_{i=1} e^{-Y_i}.
\end{equation}

\noindent \textbf{What is the actual microscopic object?} 
Although Hori and Vafa derived the correct superpotential, the mismatch in the zero-mode counting still needs to be understood. 
One clue from their observation is that the vortex is not the minimal object in the associated statistical field theory. 
Let us recall that the vortex equations in 2D GLSMs~\cite{Witten:1993yc} are
\begin{align}\label{2d:VEQ1}
\sigma &= 0, \\\label{2d:VEQ2}
D_{\bar z}\phi_i &= 0,\\\label{2d:VEQ3}
F_{12} &= e^2\left(\sum_i w_i|\phi_i|^2 - r\right).
\end{align}
The degree $k$ of the vortex is given by the topological charge
\begin{equation}\nonumber 
k = -\frac{1}{2\pi} \int F_{12}\, d^2x.
\end{equation}
Before studying the general situation, we consider two special cases:
\begin{itemize}
    \item For the charge-$N$ Abelian Higgs model, the story is simple. The effective theory decomposes into $N$ copies~\cite{Pantev:2005zs}. For each sector, the vortex indeed has two bosonic zero modes and two fermionic zero modes, and thus contributes a term
\begin{equation}
\int d^2x\, d^2\tilde{\theta}\ e^{-Y}.
\end{equation}
The factors $d^2x$ and $d^2\tilde{\theta}$ in the measure arise from the two bosonic zero modes and two fermionic zero modes, respectively.

    \item For the GLSM of $\mathbb{CP}^{N-1}$, the situation is more subtle. Unlike the previous case, there is no $\mathbb{Z}_N$ one-form symmetry that reduces the number of zero modes. However, this is analogous to many semiclassical situations in which a single topological sector admits several distinct contributing channels. The vortex is still a single one-vortex configuration, but its contribution decomposes into $N$ semiclassical channels, labeled by the matter component that degenerates at the core, leading to the sum $\sum_{i=1}^{N} e^{-Y_i}$. This may appear somewhat ad hoc. A useful way to understand the label $i$ is as follows. It does not mean that all fields vanish simultaneously, nor that each phase winds independently within a single lump configuration. Rather, $i$ labels the $i$-th homogeneous coordinate of $\mathbb{CP}^{N-1}$. Geometrically, this corresponds to the $N$ toric divisors
\[
D_i = \{\phi_i = 0\}, \qquad i = 1, \dots, N.
\]
Each divisor defines a basic instanton channel in which the map approaches $D_i$. In such a configuration, only the $i$-th field develops a zero while the others remain nonvanishing. Consequently, the fermionic zero modes split into $N$ independent pairs, each associated with one divisor $D_i$. Each channel therefore carries precisely two fermionic zero modes, which is the correct number required for a contribution to the superpotential.
\end{itemize}

The general case can be understood as a combination of the two special cases discussed above. A similar idea was recently applied to bosonic theories in~\cite{Nguyen:2023rww}, where the relevant statistical objects were described as a refined instanton gas rather than 2D vortices.  To match the nonlinear sigma model (NLSM) results, imaginary charges were assigned to the refined instantons. While this proposal may appear novel, it successfully accounts for a wide range of phenomena. In contrast to earlier results~\cite{Witten:1978bc}, the refined instanton analysis correctly captures the IR physics, eliminating the discrepancy between instanton and large-$N$ analyses.

\subsection{The IR Effective Theory}\label{GL:ets}
All these theories flow to the same IR effective theory once their dynamical parameters are matched, reflecting a universal TQFT structure in the low-energy description. This effective theory is described by a twisted effective superpotential on the $\Sigma$-space, as in equation~(\ref{2d:TESP}) :
\begin{equation}
    W_{\rm eff}(\Sigma)=-N\Sigma\left(\log\left(\frac{\Sigma}{\Lambda_{2d}}\right)-1\right).
\end{equation}
The vacuum equation $\sigma^N = \Lambda_{2d}^N$ admits $N$ solutions
\[
\sigma_c = \Lambda_{2d} e^{\frac{2\pi i j}{N}}, \quad j = 0, \dots, N-1.
\]
Thus, the axial $U(1)$ R-symmetry is broken to $\mathbb{Z}_{2N}$. 
Moreover, this equivalence extends to BPS domain walls. 
In the effective theory, domain walls (solitons in 2D) interpolate between vacua, with central charge
\[
Z_{jk} \propto W_{\rm eff}(\sigma_j) - W_{\rm eff}(\sigma_k).
\]
So, the twisted effective superpotential in the IR determines the possible BPS kink sectors connecting different vacua as well. However, the existence of these sectors does not imply that every UV GLSM contains fundamental charged matter realizing them as dynamical excitations \cite{Gu:2025gtb}. The effective theory determines the spectrum of allowed BPS charges and tensions, while the UV completion selects which of these correspond to genuine dynamical states. 
In the mirror description, these domain walls are realized as non-local configurations in the $Y$-field space.

In summary, what we have learned in 2D is as follows:
\begin{itemize}
    \item Several different 2D theories can share the same vacuum structure and BPS domain walls.
    \item Integrating out the heavy matter fields that excite these domain walls generates the same effective twisted superpotential in the IR for these different theories.
    \item Not every 2D theory admits instantons with the correct number of zero modes required to generate a superpotential term.
\end{itemize}
These observations may shed light on the origin of the IR superpotential in 4D $\mathcal{N}=1$ super Yang--Mills theory, which we will study in the following sections.

\section{Basics of 4D $\mathcal{N}=1$ SYM}\label{4d:SYM}
Before discussing the new results, we briefly review some well-known aspects of 4D $\mathcal{N}=1$ SYM. 
We focus on the $SU(N)$ gauge theory and introduce the complexified gauge coupling
\begin{equation}\label{4d:CG}
\tau
=
\frac{\Theta}{2\pi}
+
\frac{4\pi i}{g_{\rm YM}^2},
\end{equation}
where $\Theta$ is the topological theta angle and $g_{\rm YM}$ is the Yang--Mills coupling. 
The fields in the $\mathcal{N}=1$ vector multiplet are
\[
(A_\mu,\, \lambda_\alpha,\, D),
\]
which transform in the adjoint representation of the gauge group.

The ultraviolet (UV) action of four-dimensional $\mathcal{N}=1$ supersymmetric Yang--Mills theory is given by
\begin{equation} \label{4dSYM:UVA}
S_{\mathrm{gauge}}
=
\frac{1}{32\pi i}
\int d^4x\, d^2\theta\;
\tau\, \operatorname{Tr}\!\bigl(W^\alpha W_\alpha\bigr)
-
\frac{1}{32\pi i}
\int d^4x\, d^2\bar\theta\;
\bar\tau\, \operatorname{Tr}\!\bigl(\bar W_{\dot\alpha}\bar W^{\dot\alpha}\bigr)\,.
\end{equation}
The field strength superfield $W_\alpha$ is defined by
\begin{equation} \label{4dSf:CF}
W_\alpha
= -\frac{1}{4}\bar{D}^2
\left(
e^{-V} D_\alpha e^{V}
\right),
\end{equation}
where $V$ is the vector superfield. 
In the Wess--Zumino gauge, the component expansion of $W_\alpha$ is
\begin{equation}\label{4dSf:CF2}
W_\alpha
=
-i\lambda_\alpha
+\theta_\alpha D
-\frac{i}{2}(\sigma^{\mu\nu}\theta)_\alpha F_{\mu\nu}
+\theta^2 (\sigma^\mu D_\mu \bar{\lambda})_\alpha .
\end{equation}
We can then expand the superspace action to obtain the component action
\begin{equation} \label{4dSYM:UVCF}
S=
\int d^4x\left[
-\frac{1}{4g_{\rm YM}^2}\,\operatorname{Tr}(F_{\mu\nu}F^{\mu\nu})
+\frac{\Theta}{32\pi^2}\,\operatorname{Tr}(F_{\mu\nu}\widetilde F^{\mu\nu})
+\frac{i}{g_{\rm YM}^2}\,\operatorname{Tr}(\bar\lambda \bar\sigma^\mu D_\mu \lambda)
+\frac{1}{2g_{\rm YM}^2}\,\operatorname{Tr}(D^2)
\right],
\end{equation}
where the gauge field strength is
\begin{equation} 
F_{\mu\nu}
=
\partial_\mu A_\nu
-
\partial_\nu A_\mu
+
i[A_\mu,A_\nu],
\end{equation}
and the covariant derivative acting on the gaugino is
\begin{equation} 
D_\mu \lambda
=
\partial_\mu \lambda
+
i[A_\mu,\lambda].
\end{equation}
The perturbative quantum correction to the parameter $\tau$ is similar to that of the 2D complexified FI parameter $t$. 
More precisely, in terms of the holomorphic coupling $\tau$, the perturbative running is one-loop exact:
\begin{equation}
\mu \frac{d\tau}{d\mu}
=
\frac{3N}{2\pi} i .
\end{equation}

Equivalently,
\begin{equation}
\mu \frac{d}{d\mu}
\left(\frac{8\pi^2}{g_{\rm YM}^2}\right)
=
3N .
\end{equation}
Integrating the renormalization group equation leads to the definition of the RG-invariant dynamical scale $\Lambda$:
\begin{equation} 
\Lambda^{3N}
=
\mu^{3N} e^{2\pi i \tau(\mu)} .
\end{equation}

Equivalently,
\begin{equation}
\Lambda^{3N}
=
\mu^{3N}
\exp\!\left(
-\frac{8\pi^2}{g_{\rm YM}^2(\mu)} + i\Theta
\right).
\end{equation}
Since the beta function is negative at weak coupling, the theory is asymptotically free in the UV and becomes strongly coupled in the IR.

\subsection{IR Strong-Coupling Superpotential}
In the infrared, pure four-dimensional $\mathcal{N}=1$ supersymmetric Yang--Mills theory with gauge group $SU(N)$ is described by the VY superpotential in terms of the glueball superfield $S$, which is expressed in terms of the UV fields as
\begin{equation}\label{4d:Co}
S
=
-\frac{1}{32\pi^2}\operatorname{Tr} W^\alpha W_\alpha .
\end{equation}
Its component expansion is
\begin{equation}\label{4d:GF}
S
\sim
\operatorname{Tr}(\lambda\lambda)
+
\theta\,\operatorname{Tr}(F\lambda)
+
\theta^2\!\left(
\operatorname{Tr} F^2
+
i\,\operatorname{Tr} F\widetilde F
+
\cdots
\right),
\end{equation}
which is the component expansion of a gauge-invariant composite chiral operator. The effective VY superpotential is
\begin{equation}\label{4d:VY}
W_{\rm VY}(S)
=
N\,S\left(1-\log\frac{S}{\Lambda^3}\right).
\end{equation}

The supersymmetric vacua are determined by
\begin{equation}
\frac{\partial W_{\rm VY}}{\partial S}=0,
\end{equation}
which implies
\begin{equation}
\langle S\rangle
=
\Lambda^3\, e^{2\pi i k/N},
\qquad
k=0,1,\dots,N-1.
\end{equation}

Thus, the theory has $N$ supersymmetric vacua and exhibits gaugino condensation,
\begin{equation}
\langle \Tr \lambda^\alpha \lambda_\alpha \rangle \neq 0.
\end{equation}
This corresponds to the spontaneous breaking
\begin{equation}
\mathbb Z_{2N}\to \mathbb Z_2 .
\end{equation}
One can readily observe that the vacuum structure is very similar to that of the 2D cases discussed in Section~\ref{Les:GLSM}.

\paragraph{Infrared three-form multiplet}In the microscopic theory, one has the composite chiral operator
\begin{equation}
S_{\rm UV}
=
-\frac{1}{32\pi^2}\operatorname{Tr} W^\alpha W_\alpha .
\end{equation}
This operator captures the relevant holomorphic sector of pure 4D $\mathcal{N}=1$ $SU(N)$
super Yang--Mills theory. 
In the far infrared, however, the theory is gapped. 
Therefore, the deep IR bulk is not described in terms of ordinary propagating degrees of freedom. 
Instead, what survives are the low-energy holomorphic data: the gaugino condensate, the $N$ vacua, the $\theta$-dependence, and the central charges of BPS domain walls.

For this reason, one replaces the microscopic composite description with an effective infrared representation in terms of a special chiral superfield $S_{\rm IR}$ associated with a three-form multiplet. 
Its superspace expansion is
\begin{equation} \label{4d:IRS}
S_{\rm IR}(y,\theta)
=
s(y)
+
\sqrt{2}\,\theta \chi(y)
+
\theta^2 \mathcal{F}(y),
\qquad
y^\mu = x^\mu + i\theta\sigma^\mu\bar\theta .
\end{equation}

For a three-form multiplet, the highest component is not a generic complex auxiliary field.
Rather, it takes the form
\begin{equation} 
\mathcal F
=
F_{\rm IR}
+
i\,{}^\ast F_4,
\end{equation}
where $F_{\rm IR}$ is a real auxiliary scalar and
\begin{equation}
F_4=dC_3
\end{equation}
is the four-form field strength of a three-form gauge field $C_3$.

Thus one may write
\begin{equation}\label{4d:IRS2}
S_{\rm IR}(y,\theta)
=
s(y)
+
\sqrt{2}\,\theta \chi(y)
+
\theta^2\bigl(F_{\rm IR}+i\,{}^\ast F_4\bigr).
\end{equation}
This three-form realization does not reproduce the microscopic operator
$-\frac{1}{32\pi^2}\operatorname{Tr} W^\alpha W_\alpha$ component by component. 
Rather, it captures the same infrared holomorphic sector while being compatible with the mass gap: in four dimensions, a three-form carries no local propagating bulk degrees of freedom, but its four-form field strength correctly encodes the vacuum branches and domain-wall charges. See

Thus, the UV-to-IR transition
\begin{equation}
-\frac{1}{32\pi^2}\operatorname{Tr} W^\alpha W_\alpha
\quad \rightsquigarrow \quad
S_{\rm IR}
\end{equation}
should be understood as a holomorphic matching of the low-energy chiral sector, rather than a microscopic component-by-component identification.

\subsection{BPS Domain Walls in 4D $\mathcal{N}=1$ SYM}\label{4D:DSY}

As in the discussion in Section~\ref{GL:ets} for 2D, one can also define BPS domain walls in 4D~\cite{Dvali:1996xe}. 
Pure four-dimensional $\mathcal{N}=1$ $SU(N)$ supersymmetric Yang--Mills theory has $N$ discrete vacua, which admit domain walls interpolating between them. 
A domain wall preserving one-half of the supersymmetry is called a BPS domain wall. 
A wall connecting the $j$-th vacuum and the $(j+k)$-th vacuum is referred to as a $k$-wall. 
Its tension is proportional to
\begin{equation} 
T_k
\propto
N \Lambda^3 \sin\!\left(\frac{\pi k}{N}\right),
\end{equation}
up to convention-dependent normalization factors. 

Motivated by the two-dimensional analysis, one may ask whether the effective VY superpotential can be derived by integrating out the degrees of freedom that generate the domain walls. This question cannot be addressed within pure super Yang--Mills theory alone. 
We will revisit this issue in the next section.
\section{Domain Walls as Dynamical Fields: Field-Theoretic Description and Linear--Chiral Duality}\label{4d:DDWD}

In pure $\mathcal{N}=1$ supersymmetric Yang--Mills theory, a single instanton cannot directly generate a superpotential term due to the mismatch in fermionic zero modes. We now elaborate on this point. According to the Atiyah--Singer index theorem, a Weyl fermion in a representation $\mathcal{R}$ of the gauge group has
\[
n_{\text{zero}} = 2\,T(\mathcal{R})\,k
\]
zero modes in an instanton background with topological charge $k$, where $T(\mathcal{R})$ is the quadratic index of the representation. 
For $SU(N)$ gauge theory, the gaugino transforms in the adjoint representation with $T(\mathrm{adj}) = N$. 
Therefore, a one-instanton configuration ($k=1$) carries
\[
2N
\]
gaugino zero modes, which exceeds the two zero modes required to generate a superpotential term. A superpotential term arises from a $d^2\theta$ superspace integral and can absorb only two fermionic zero modes. 
In contrast, an instanton background contains $2N$ gaugino zero modes, preventing a direct contribution to the superpotential. 
Instead, it generates multi-fermion interactions of the schematic form
\[
(\lambda\lambda)^N .
\]
Consequently, the Veneziano--Yankielowicz superpotential in 4D $\mathcal{N}=1$ SYM should not be interpreted as being generated directly by instantons. 
Rather, it provides an effective description of the strong-coupling dynamics of the theory, constrained by holomorphy, anomalies, and the discrete chiral symmetry.\footnote{It has been commonly believed that this conclusion is specific to four-dimensional instantons on $\mathbb{R}^4$. 
In more general settings, however, non-perturbative configurations can generate a superpotential once the number of fermionic zero modes is reduced to two. 
For example, as mentioned in the introduction, upon compactification on $\mathbb{R}^3 \times S^1$, a four-dimensional instanton can fractionalize into monopole-instantons, each of which carries only two fermionic zero modes and can therefore contribute directly to the superpotential. 
Likewise, in theories with additional matter fields, deformations, background fields, or interaction terms that lift the extra zero modes, instanton effects can generate superpotential terms.}

However, we revisit this statement by providing a direct derivation within our extended theory.
This theory shares the same vacuum structure and BPS domain walls, and therefore lies in the same universality class as 4D $\mathcal{N}=1$ SYM.

\subsection{Quantum Fields for Domain Walls in 4D $\mathcal{N}=1$ SYM}\label{4dSYM:QFD}
Our analysis in two dimensions shows that elementary matter fields are responsible for dynamical domain walls. 
In four dimensions, dynamical domain walls are typically described in terms of their worldvolume theories, since they are extended objects. 
However, this perspective is not useful for our purposes. 

This raises the question: what are the candidate elementary fields that can excite dynamical domain walls in 4D $\mathcal{N}=1$ SYM? 
At first sight, this may appear to be an ill-posed question, since the fundamental excitations are extended objects rather than point-like particles. 
However, this conclusion is premature. 

For example, in gauge theory one can define Wilson loop operators, which describe extended objects in terms of fundamental gauge fields. 
By analogy, describing domain walls as dynamical objects would require a higher-form gauge field. 
However, such a field is not present in pure 4D $\mathcal{N}=1$ SYM. 

This suggests that one should embed pure 4D $\mathcal{N}=1$ SYM into a larger theory that shares the same vacuum structure and BPS domain walls. 
Based on our experience in two dimensions, this is indeed possible. 
Let us therefore recall what plays the role of the elementary field for domain walls in 2D:
\begin{equation} \nonumber
\rho\, \exp\!\left(i \int_{\gamma} (d\varphi + Q A) \right) + \cdots,
\end{equation}
where ``$\cdots$'' denotes the supersymmetric completion of the operator. 
By analogy, we expect that the following operator
\begin{equation} \label{4dSYM:DF}
\rho\, \exp\!\left(
i\int_{\Sigma=\partial\mathcal{M}} B_2
+ iQ\int_{\mathcal{M}} A_{3}
\right)
+ \cdots
\end{equation}
can create (or excite) a domain wall in 4D $\mathcal{N}=1$ SYM. Another hint comes from the axion duality. Perturbatively, the axion enjoys a continuous shift symmetry, known as the Peccei--Quinn symmetry.
It is useful to understand how this symmetry is realized in the dual $(d-2)$-form description. We will see momentarily that a massless $(d-2)$-form gauge field in $d$ spacetime dimensions carries a single physical propagating degree of freedom and is therefore dual to a scalar field. In the dual description, one may formally write a shift transformation
\[
B_2 \;\to\; B_2 + \omega_2, 
\qquad d\omega_2 = 0,
\]
i.e. a shift by a closed two-form. 
However, counting such $\omega_2$ (even modulo gauge transformations) does not directly determine the number of independent global symmetry parameters.

Instead, the correct characterization is in terms of the conserved charge. 
In both the axion and dual two-form descriptions, there is a single conserved generator,
\[
\mathcal{Q} \;=\; \int_{\Sigma_3} *\, da \;=\; \int_{\Sigma_3} dB_2,
\]
where the equality follows from the duality relation
\[
dB_2 \;\propto \;  * da,
\]
up to normalization. 
Therefore, the dual theory possesses a single independent continuous symmetry parameter, matching the Peccei--Quinn shift symmetry of the axion.

In this language, domain walls arise as configurations interpolating between different branches associated with the periodicity of the axion. 
In the dual formulation, the phase variable is replaced by higher-form gauge fields, and the coupling to extended objects such as domain walls is naturally encoded by higher-dimensional Wilson-type operators (\ref{4dSYM:DF}). 

The corresponding elementary fields can be described in terms of
\begin{equation} \nonumber
\left(\rho,\; B_{\mu\nu}\right).
\end{equation}
It is instructive to represent them in different coordinate systems; in this paper, however, we work in polar coordinates. 
From the index structure of the ``phase'' and the three-form gauge field $A_{3}$ in the above field decomposition, one observes that the relevant gauge symmetry is a two-form gauge symmetry. 
The corresponding charged matter fields and gauge field are $\left(\rho,\; B_{\mu\nu}\right)$ and $A_{3}$, respectively. 

Although $\left(\rho,\; B_{\mu\nu}\right)$ is referred to as a ``matter'' field in our setup, it is important to emphasize that $B_{\mu\nu}$ itself is a gauge field. 
As a result, the fundamental excitations are extended objects rather than point-like particles. 
Before writing down a Lagrangian, let us first examine the field content by counting the degrees of freedom.
\paragraph{Degrees of freedom of $p$-form fields in $d$ dimensions}

For a field of $p$-form in a spacetime of $d$, the number of physical propagating degrees of freedom is

\begin{itemize}
\item \textbf{Massless $p$-form gauge field:}
\[
N_{\text{massless}}(d,p) = \binom{d-2}{p}.
\]

\item \textbf{Massive $p$-form field:}
\[
N_{\text{massive}}(d,p) = \binom{d-1}{p}.
\]
\end{itemize}

These formulas arise because the little group of a particle in $d$ dimensions is $SO(d-2)$ for massless particles and $SO(d-1)$ for massive particles. The physical polarizations therefore transform as a $p$-form in the corresponding transverse space, whose dimension is given by the binomial coefficients above.

For $d=4$:

\begin{itemize}

\item \textbf{2-form gauge field ($p=2$):}
\[
N_{\text{massless}}(4,2) = \binom{2}{2} = 1,
\qquad
N_{\text{massive}}(4,2) = \binom{3}{2} = 3.
\]

Thus a massless 2-form gauge field in four dimensions carries one propagating degree of freedom.

\item \textbf{3-form gauge field ($p=3$):}
\[
N_{\text{massless}}(4,3) = \binom{2}{3} = 0,
\qquad
N_{\text{massive}}(4,3) = \binom{3}{3} = 1.
\]

Therefore a massless 3-form gauge field in four dimensions has no local propagating degrees of freedom, while a massive 3-form has one.

\end{itemize}

\paragraph{3-form Higgs/St\"uckelberg mechanism in four dimensions}

In four dimensions, a 3-form gauge field $A_{3}$ can become massive by eating a 2-form gauge field $B_2$ through the gauge-invariant combination
\[
\mathcal{H}_3 = dB_2 + Q A_{3}.
\]
A corresponding Lagrangian is
\[
-\frac{1}{2e_{4d}^2} |dA_3|^2
-|dB_2 + Q A_3|^2.
\]
This is invariant under the gauge transformations
\[
A_3 \to A_3 - d\Lambda_2,
\qquad
B_2 \to B_2 + Q\Lambda_2 + d\Lambda_1,
\]
where $\Lambda_2$ is a 2-form gauge parameter and $\Lambda_1$ is a 1-form gauge parameter.
\[
\mathcal{L}
=
-\frac{1}{2e_{4d}^2} |dA_3|^2
-\rho^2 |dB_2 + Q A_3|^2-(\partial_{\mu}\rho)^2.
\]
The degree-of-freedom counting is
\[
1 \quad (B_2\ \text{massless})
\;+\;
0 \quad (A_3\ \text{massless})
\;\longrightarrow\;
1 \quad (A_3\ \text{massive}).
\]
Hence, in four dimensions, the St\"uckelberg coupling $dB_2 + Q A_3$ realizes a Higgs mechanism for a three-form gauge field. 
As in the 2D case, we include a radial mode. 

Since we are primarily interested in BPS domain walls in 4D $\mathcal{N}=1$ theories, these bosonic fields must be embedded into appropriate supermultiplets, which we discuss in the next section.

\subsection{A Lagrangian for BPS Domain Walls}\label{4dSYM:LDD}
In the previous section, we proposed a bosonic field description of domain walls with the Lagrangian
\[
\mathcal{L}
=
-\frac{1}{2e^2} |dA_3|^2
-\rho^2 |dB_2 + Q A_3|^2
-(\partial_{\mu}\rho)^2.
\]
To describe all fundamental domain walls in 4D $\mathcal{N}=1$ $SU(N)$ SYM, one would need to introduce $N$ $B$-fields with unit charges. 
However, as discussed in Section~\ref{Les:GLSM} for the 2D case, different theories may flow to the same IR theory. 
Motivated by this, we consider a simplified setup with a single $B$-field of charge $N$. 

In a more general situation, additional massless fields would be present, for which the Higgs mechanism alone is insufficient to generate masses. 
Nevertheless, we expect these fields to become massive, which is closely related to the problem of the mass gap~\cite{Dierigl:2014xta}. 
We further expect that non-perturbative effects provide such a mechanism, and we will comment on this when deriving the superpotential. 

To construct a supersymmetric Lagrangian and analyze its exact properties, it is natural to organize the fields into supersymmetry multiplets.

\paragraph{Linear multiplet description of $(\rho, B_2)$ in 4D $\mathcal{N}=1$.}

We promote the bosonic fields $(\rho, B_2)$ to a four-dimensional $\mathcal{N}=1$ supermultiplet by embedding them into a real linear superfield $L$ in flat superspace, defined by
\begin{equation}\label{4D:LM}
L = L^\dagger, \qquad D^2 L = \bar D^2 L = 0 \, .
\end{equation}
The component fields are obtained via superspace projections,
\begin{equation}\label{4D:LMC}
C = L\big|, \qquad
\chi_\alpha = - i\, D_\alpha L\big|, \qquad
H_{\alpha\dot\alpha} = -\frac12 [D_\alpha, \bar D_{\dot\alpha}] L\big| ,
\end{equation}
where $|$ denotes the evaluation in $\theta = \bar\theta = 0$. 
The linear constraint implies that the vector $H_\mu$ is divergence-free, $\partial^\mu H_\mu = 0$, and can therefore be written locally as the Hodge dual of a three-form field strength,
\begin{equation}
H^\mu = \frac{1}{3!}\, \epsilon^{\mu\nu\rho\sigma} H_{\nu\rho\sigma}, 
\qquad H_3 = d B_2 \, .
\end{equation}
Thus, the real linear multiplet describes the field content $L \sim (C, \chi_\alpha, H_{\mu\nu\rho})$, and the two-form gauge field $B_{\mu\nu}$ enters only through its gauge-invariant field strength $H_3$. 
The superfield expansion takes the form
\begin{equation} 
\begin{aligned}
L(x,\theta,\bar\theta)
={}& C
+ i\theta\chi - i\bar\theta\bar\chi
+ \theta\sigma^\mu\bar\theta\, H_\mu \\
&\quad
-\frac12 \theta^2 \bar\theta\, \bar\sigma^\mu \partial_\mu \chi
-\frac12 \bar\theta^2 \theta\, \sigma^\mu \partial_\mu \bar\chi
-\frac14 \theta^2 \bar\theta^2 \Box C \, ,
\end{aligned}
\end{equation}
with $H^\mu = \tfrac{1}{3!}\epsilon^{\mu\nu\rho\sigma} H_{\nu\rho\sigma}$. 
One may identify the following field map:
\begin{equation} 
C := L| = 2\log \rho \, ,
\qquad \rho = e^{C/2} .
\end{equation}

The simplest supersymmetric kinetic term is given by the D-term
\begin{equation} 
\mathcal L = \int d^4\theta\, e^{L} \, ,
\end{equation}
which reproduces the bosonic Lagrangian
\begin{equation}\label{4d:DBL}
\mathcal{L} 
=
- (\partial_\mu \rho)(\partial^\mu \rho)
- \frac{1}{6}\,\rho^2 H_{\mu\nu\rho} H^{\mu\nu\rho},
\qquad
H_3 = dB_2.
\end{equation}
However, the fermionic sector (i.e., $\chi$) is not free, in contrast to the situation in two-dimensional GLSMs.

\paragraph{Three-form multiplet and the gauge field $A_3$.}

The three-form gauge field $A_3$ is naturally embedded into a four-dimensional $\mathcal{N}=1$ three-form multiplet~\cite{Gates:1980ay}. 
In flat superspace, this multiplet may be described in terms of a real prepotential $U = U^\dagger$, from which one constructs the chiral field strength
\begin{equation}\label{4d:TFM}
\mathcal{S} = -\frac14 \bar D^2 U, \qquad \bar D_{\dot\alpha}\mathcal{S} = 0 \, .
\end{equation}
The prepotential is defined modulo the gauge transformation
\begin{equation}
U \;\to\; U + \tilde{L} \, ,
\end{equation}
where $\tilde{L}$ is a real linear superfield satisfying
\begin{equation}
\tilde{L} = \tilde{L}^\dagger, \qquad D^2\tilde{L} = \bar D^2\tilde{L} = 0 \, .
\end{equation}
This ensures that $\mathcal{S}$ is gauge invariant. 
In components, the chiral superfield can be expanded as
\begin{equation} 
\mathcal{S} = s + \sqrt{2}\,\theta\eta + \theta^2 \mathcal{F}
+ i \theta \sigma^\mu \bar\theta\, \partial_\mu s 
-\frac{i}{\sqrt{2}} \theta^2 \partial_\mu \eta \sigma^\mu \bar\theta
-\frac14 \theta^2 \bar\theta^2 \Box s  \, ,
\end{equation}
where $s = \mathcal{S}|$ is a complex scalar, $\eta_\alpha = \frac{1}{\sqrt{2}} D_\alpha \mathcal{S}|$ is a Weyl fermion, and the highest component $\mathcal{F}$ contains the four-form field strength of the three-form gauge field,
\begin{equation}
\mathcal{F} = F + i\, {*F_4}, \qquad F_4 = dA_3 \, .
\end{equation}
Thus, the three-form multiplet can be viewed as a variant of a chiral multiplet in which the highest component is not an unconstrained auxiliary complex scalar, but instead contains the Hodge dual of the gauge-invariant four-form field strength associated with $A_3$. 
Its superspace Lagrangian is
\[
\mathcal{L}_{\text{gauge}}
=
\frac{1}{2e^2}\int d^4\theta\, \bar{\mathcal{S}}\, \mathcal{S} \, .
\]

\paragraph{Three-form realization of the Fayet--Iliopoulos term.}
In analogy with the ordinary Fayet--Iliopoulos (FI) coupling, one may consider a linear coupling to the three-form prepotential $U=U^\dagger$,
\[
\mathcal{L}_{\text{FI}}
=
-\xi \int d^4\theta\, U .
\]
In terms of component fields, it is 
\begin{equation}
-\xi F \, .
\end{equation}

With the above ingredients, we can now write the following Lagrangian for the domain wall in 4D $\mathcal{N}=1$ SYM.

\paragraph{A Lagrangian.}

To supersymmetrize the bosonic combination
\begin{equation}\label{4d:TFGI}
\mathcal{H}_3 = dB_2 + N A_3 \, ,
\end{equation}
we introduce a \emph{chiral spinor prepotential} $\Sigma_\alpha$ for the two-form $B_2$, satisfying
\begin{equation}\label{4d:CSP}
\bar{D}_{\dot{\alpha}} \Sigma_{\beta} = 0.
\end{equation}
In chiral coordinates
\begin{equation}
y^m = x^m + i \theta \sigma^m \bar{\theta} ,
\end{equation}
the superfield admits the expansion
\begin{equation}
\Sigma_{\alpha}(y,\theta)
=
\zeta_{\alpha}(y)
+ \theta_{\alpha} M(y)
+ (\sigma^{mn}\theta)_{\alpha}\, B_{mn}(y)
+ \theta^2 \rho_{\alpha}(y) ,
\end{equation}
where $\zeta_{\alpha}$ and $\rho_{\alpha}$ are Weyl fermions, $M$ is a complex scalar, and $B_{mn}=-B_{nm}$ is an antisymmetric tensor. The latter is encoded in $(\sigma^{mn}\theta)_\alpha$ and gives rise, through the gauge-invariant combination defining $L$, to the physical real two-form gauge potential.

The prepotential $\Sigma_{\alpha}$ is defined up to the gauge transformation
\begin{equation} 
\delta \Sigma_{\alpha}
=
-\frac{1}{4}\,\bar{D}^2 D_{\alpha} \Theta ,
\qquad
\Theta^{\dagger} = \Theta ,
\end{equation}
which preserves chirality. The corresponding gauge-invariant real superfield is
\begin{equation} \label{4d:GIRS}
L
=
\frac{1}{2i}
\left(
D^{\alpha} \Sigma_{\alpha}
-
\bar{D}_{\dot{\alpha}} \bar{\Sigma}^{\dot{\alpha}}
\right) ,
\end{equation}
which satisfies the linear constraints
\begin{equation} 
D^2 L = 0,
\qquad
\bar{D}^2 L = 0 ,
\end{equation}
and therefore defines a real linear multiplet. This differs from some standard conventions by an overall factor of $i$ in the definition of the chiral spinor prepotential.

Using the gauge freedom in $\Theta$, one can choose a Wess--Zumino-type gauge in which the lowest spinor and scalar components are removed,
\begin{equation} 
\zeta_{\alpha} = 0, \qquad M = 0, 
\end{equation}
leaving only the two-form gauge potential and its fermionic partner:
\begin{equation} 
\Sigma_{\alpha}^{\mathrm{WZ}}
=
(\sigma^{mn}\theta)_{\alpha} B_{mn}
+
\theta^2 \rho_{\alpha} .
\end{equation}

The residual gauge symmetry acts on the two-form as
\begin{equation} \label{4d:RGT}
\delta B_{mn}
=
2 \partial_{[m} \Lambda_{n]} ,
\end{equation}
which is the standard gauge transformation of a two-form field.

In the presence of the three-form $A_3$, this symmetry is extended to a
Stückelberg-type transformation,
\begin{equation}\nonumber
\delta B_2 = d\Lambda_1 + N \Lambda_2,
\qquad
\delta A_3 = -d\Lambda_2,
\end{equation}
such that the combination
\begin{equation}
\mathcal{H}_3 = dB_2 + N A_3
\end{equation}
is gauge invariant.

We have already introduced a real prepotential $U=U^\dagger$ for the three-form $A_3$. We now construct the Stückelberg-covariant real superfield
\begin{equation}\label{4d:GCRS}
\mathbb L = L + N U \, .
\end{equation}
Its lowest component defines the radial scalar,
\begin{equation} 
C := \mathbb L| = 2\log \rho \, ,
\qquad \rho = e^{C/2} .
\end{equation}
Its $\theta\bar\theta$ component contains the Hodge dual of the gauge-invariant field strength $\mathcal{H}_3 = dB_2 + N A_3$. 
The superfield $\mathbb L$ obeys the deformed linear constraint
\begin{equation}\label{4d:DLC}
-\frac14 \bar D^2 \mathbb L = N\,\mathcal{S} \, ,
\end{equation}
which corresponds to the Bianchi identity
\begin{equation} \nonumber
d\mathcal{H}_3 = N F_4, \qquad F_4 = dA_3 \, .
\end{equation}

The Fayet--Iliopoulos-type coupling for the three-form multiplet can be written either as a D-term linear in $U$ or as an F-term linear in $\mathcal{S}=-\frac14\bar D^2U$,
\begin{equation} \nonumber
-\xi \int d^4\theta\, U
\;=\;
-\frac{\xi}{2} \left(\int d^2\theta\, \mathcal{S}+\text{h.c.}\right),
\end{equation}
up to boundary terms. In particular, it can be written as a linear superpotential \cite{Cribiori:2018jjh},
\begin{equation} 
W(\mathcal{S}) = \tilde{\tau}\,\mathcal{S},
\end{equation}
with complex parameter $\tilde{\tau}=\xi-i\tilde{\Theta}$. Since the highest component of $\mathcal{S}$ contains
\begin{equation} 
F + i\, {*F_4}, \qquad F_4 = dA_3,
\end{equation}
the real part of $\tilde{\tau}$ couples to the auxiliary field $F$, while the imaginary part couples to the four-form flux and plays the role of the theta angle.

A superspace action reproducing a bosonic Higgs-type structure with canonical radial kinetic term is
\begin{equation}\label{4d:DWSA}
S
=
\int d^4x\, d^4\theta\, \left(e^{\mathbb L}+\frac{1}{2e^2} \bar{\mathcal{S}}\, \mathcal{S}\right)
-\frac{1}{2} 
\left[
\int d^4x\, d^2\theta\, W(\mathcal S)
+\text{h.c.}
\right].
\end{equation}

Since
\begin{equation} 
K''(C)=e^{C}=\rho^2,
\end{equation}
the bosonic sector takes the form
\begin{equation} \label{4d:DABA}
\mathcal L_{\rm bos}
=
-(\partial_\mu\rho)(\partial^\mu\rho)
-\frac{1}{6}\,\rho^2 \mathcal{H}_{\mu\nu\rho}\mathcal{H}^{\mu\nu\rho}
-\frac{1}{2e^2} |dA_3|^2
+\mathcal L_W,
\qquad
\mathcal{H}_3 = dB_2 + N A_3 ,
\end{equation}
where $\mathcal L_W$ denotes the contribution from $W(\mathcal S)$. The precise normalization of the $\mathcal H_3^2$ term depends on the convention
used for the dual vector $\mathcal H_\mu$.
A complete component expansion is presented in Appendix~\ref{4d:DCL}.

\paragraph{Supersymmetry transformation.}

The supersymmetry transformations in the $(C,\chi)$ variables can be written in a form analogous to that of a non-linear sigma model:
\begin{align}
\delta C &= i(\epsilon \chi - \bar\epsilon \bar\chi), \\[4pt]
\delta \chi_\alpha &=
(\sigma^\mu \bar\epsilon)_\alpha (\partial_\mu C + i \mathcal{H}_\mu)
+ 2N \epsilon_\alpha s
- \Gamma(C)\,\delta C\,\chi_\alpha, \\[4pt]
\delta s &= \sqrt{2}\,\epsilon \eta, \\[4pt]
\delta \eta_\alpha &=
i\sqrt{2} (\sigma^\mu \bar\epsilon)_\alpha \partial_\mu s
+ \sqrt{2}\,\epsilon_\alpha (F + i *F_4), \\[4pt]
\delta (F + i *F_4) &=
i\sqrt{2}\,\bar\epsilon \bar\sigma^\mu \partial_\mu \eta.
\end{align}

It is useful to introduce the \emph{rescaled fermion}, defined as the canonically normalized field
\begin{equation}\label{4d:RSF}
\psi_\alpha
=
\sqrt{G(C)}\,\chi_\alpha
=
\frac{\rho}{2}\,\chi_\alpha .
\end{equation}
This definition is motivated as follows. 

The one-dimensional sigma-model metric is
\begin{equation}
G(C)=\frac14 K''(C)=\frac{e^C}{4}=\frac{\rho^2}{4},
\end{equation}
so that
\begin{equation}
\Gamma(C)=\frac12\,\frac{G'(C)}{G(C)}=\frac12.
\end{equation}
Accordingly, the canonically normalized fermion is given by~\eqref{4d:RSF}.
Equivalently,
\begin{equation}
\chi_\alpha = \frac{2}{\rho}\,\psi_\alpha .
\end{equation}
We can now express the supersymmetry transformations in terms of the $(\rho,\psi)$ variables. 
Using
\begin{equation}
\delta \rho
=
\frac{\rho}{2}\,\delta C ,
\end{equation}
we obtain
\begin{equation}
\delta \rho
=
i(\epsilon \psi - \bar\epsilon \bar\psi) .
\end{equation}

The rescaled fermion transforms as
\begin{equation}
\delta \psi_\alpha
=
(\sigma^\mu \bar\epsilon)_\alpha\,\partial_\mu \rho
+
\frac{i\rho}{2}\,
(\sigma^\mu \bar\epsilon)_\alpha\,\mathcal{H}_\mu
+
N\rho\,\epsilon_\alpha s .
\end{equation}

As we will show in Appendix~\ref{4d:DCL}, the Lagrangian for the rescaled fermion closely resembles that of fermions in a two-dimensional GLSM. 
This is not surprising: although the 4D action appears to take the form of a non-linear sigma model, this is largely an artifact of the field redefinition, and the theory is in fact a sigma model with a flat target space.

\subsection{Linear--Chiral Duality in Supersymmetric Field Theories }\label{4d:LCD}
As in the two-dimensional case, it is often advantageous to describe the dynamics of the system in terms of dual variables. 
In two dimensions, this procedure is usually referred to as Abelian duality, while in four dimensions it is known as linear--chiral duality~\cite{Lindstrom:1983rt, Giedt:2003ap}.

\paragraph{Bosonic dualization.}
We begin with the bosonic sector. The relevant part of the action is
\begin{equation}
\mathcal L_{\rm bos}
=
-(\partial_\mu \rho)(\partial^\mu \rho)
-\frac{1}{6}\rho^2  \mathcal{H}_{\mu\nu\rho}\mathcal{H}^{\mu\nu\rho}
-\frac{1}{2e^2}\frac{1}{4!}F_{\mu\nu\rho\sigma}F^{\mu\nu\rho\sigma}
.
\end{equation}
To dualize the two-form, we introduce a first-order Lagrangian
\begin{equation} \label{4d:ABT}
\mathcal L_{\rm 1st}
=
-d\rho\wedge *d\rho
- \rho^2 \mathcal{H}_3\wedge *\mathcal{H}_3
+da\wedge \mathcal{H}_3
-Na\,F_4
-\frac{1}{2e^2}F_4\wedge *F_4
.
\end{equation}
Varying with respect to $a$ imposes the Bianchi identity
\begin{equation}
d\mathcal{H}_3=NF_4,
\end{equation}
which is solved by $\mathcal{H}_3=dB_2+NA_3$, thereby reproducing the original formulation.

On the other hand, varying with respect to $\mathcal{H}_3$ gives
\begin{equation}
da=-2\rho^2 *\mathcal{H}_3,
\end{equation}
where the relative sign follows from the graded commutativity of differential forms,
$da\wedge \delta\mathcal H_3 = -\,\delta\mathcal H_3 \wedge da$.
Substituting this relation back into the first-order action leads to the dual bosonic Lagrangian
\begin{equation}\label{4d:TDB}
\mathcal L_{\rm dual,bos}
=
-(\partial_\mu \rho)(\partial^\mu \rho)
-\frac{1}{4\rho^2}(\partial_\mu a)(\partial^\mu a)
-\frac{N}{24}\epsilon^{\mu\nu\rho\sigma}aF_{\mu\nu\rho\sigma}
-\frac{1}{2e^2}F_4\wedge *F_4.
\end{equation}

Defining
\begin{equation}
\varrho=\rho^2,
\end{equation}
the kinetic terms can be written as
\begin{equation}\label{4d:TDB2}
\mathcal L_{\rm dual,bos}
=
-\frac{1}{4\varrho}(\partial_\mu \varrho)(\partial^\mu \varrho)
-\frac{1}{4\varrho}(\partial_\mu a)(\partial^\mu a)
-\frac{N}{24}\epsilon^{\mu\nu\rho\sigma}aF_{\mu\nu\rho\sigma}
-\frac{1}{2e^2}F_4\wedge *F_4.
\end{equation}

\paragraph{Superspace duality.}
We now perform the duality in superspace, starting from the action~(\ref{4d:DWSA}). 
To implement the linear--chiral duality, we relax the constraint on $\mathbb L$ and introduce a chiral superfield $Z$ via the first-order action
\begin{equation}\label{4d:SSAD}
S_{\rm 1st}
=
\int d^4x\,d^4\theta
\left[
e^{\mathbb L}
+\frac{1}{2e^2}\bar{\mathcal S}\mathcal S
-\frac12\,\mathbb L\,(Z+\bar Z)
\right]
+\left[
\int d^4x\,d^2\theta
\left(
\frac{N}{2}Z\,\mathcal S-\frac12 W(\mathcal S)
\right)
+\text{h.c.}
\right].
\end{equation}
Varying with respect to $Z$ yields
\begin{equation}
-\frac14\bar D^2\mathbb L = N\mathcal S,
\end{equation}
which restores the original definition $\mathbb L=L+NU$. 
On the other hand, varying with respect to $\mathbb L$ gives
\begin{equation}
Z+\bar Z = 2e^{\mathbb L}.
\end{equation}

Writing the lowest component of $Z$ as
\begin{equation}
Z|=z=\varrho-i a,
\end{equation}
one finds
\begin{equation}
\varrho=e^{C}=\rho^2.
\end{equation}
Substituting back into the action leads to the dual theory
\begin{equation}\label{4d:DWDSA}
S_{\rm dual}
=
\int d^4x\,d^4\theta
\left[
-\frac12 (Z+\bar Z)\log(Z+\bar Z)
+\frac{1}{2e^2}\bar{\mathcal S}\mathcal S
\right]
-\frac12\left[
\int d^4x\,d^2\theta\,(\tilde{\tau}-NZ)\mathcal S
+\text{h.c.}
\right],
\end{equation}
where we have taken $W(\mathcal S)=\tilde{\tau}\,\mathcal S$ and dropped a term proportional to $(Z+\bar Z)$, whose full superspace integral vanishes (equivalently, it corresponds to a trivial K\"ahler transformation). 
This formulation closely parallels the two-dimensional case.

\paragraph{Symmetry interpretation.}
It is useful to make the symmetry structure underlying our construction more explicit. In the dual description, the relevant infrared degree of freedom is encoded by the chiral field
\[
Z=\varrho-i a,
\]
whose imaginary part inherits an axionic shift structure originating from the gauge invariance of the two-form field. 
At the classical level this can be viewed as a continuous shift symmetry, while in the full quantum theory it is generally reduced to a discrete subgroup once the coupling to the three-form sector and flux quantization are taken into account.

Domain walls are realized as configurations in which the scalar fields interpolate between distinct vacua. 
In the three-form formulation, the four-form field strength $F_4$ does not describe propagating degrees of freedom in four dimensions, but is instead determined algebraically by the scalar configuration (away from localized sources). 
Thus, the three-form description provides an equivalent encoding of the same infrared structure, while the wall itself is most transparently described by the interpolating scalar profile.

The coupling encoded in the combination $\mathcal H_3=dB_2+NA_3$ can be viewed as a higher-form analogue of the Stückelberg (or Higgs) mechanism, reducing the continuous structure to a discrete $\mathbb{Z}_N$ remnant. 
This reduction underlies the emergence of multiple semiclassical contributions in our framework. In the next section, we discuss the quantum corrections of the theory.

\subsection{Quantum Corrections}\label{4d:QC}

The classical Lagrangian receives both perturbative and non-perturbative quantum corrections. 
The K\"ahler potential generally acquires complicated quantum corrections, which are important for determining the full physical spectrum of the theory. 
However, if one is only interested in the BPS spectrum, it is sufficient to work in a holomorphic scheme, in which the detailed form of the K\"ahler potential can be neglected and the focus is placed entirely on the superpotential. It is well known that the Wilsonian superpotential is protected by a perturbative non-renormalization theorem. 
Nevertheless, this does not preclude modifications of the low-energy effective superpotential arising from integrating out heavy degrees of freedom or from non-perturbative effects, such as instanton contributions.

\paragraph{Scheme dependence and FI counterterms.}

Let us clarify the notion of scheme dependence in the present setup.  
In the three-form multiplet formulation, the natural local supersymmetric FI-type counterterm is
\begin{equation}
\Delta \mathcal{L}_{\rm ct}
=
\int d^4\theta \; k\, U ,
\end{equation}
where $U$ is the real three-form prepotential. 
This term is invariant under the gauge transformation
\begin{equation}
U \;\to\; U + \Lambda + \bar{\Lambda} ,
\end{equation}
with $\Lambda$ a chiral superfield. 
Indeed, the variation of the counterterm is proportional to
\begin{equation}
\int d^4\theta\, (\Lambda + \bar{\Lambda}) = 0 ,
\end{equation}
since the full superspace integral of a chiral or anti-chiral superfield vanishes (up to possible boundary terms). 
Therefore, the counterterm is gauge invariant.

At the component level, this counterterm induces a shift in the auxiliary field coupling,
\begin{equation}
\Delta \mathcal{L}_{\rm aux} = -\, k\, F ,
\end{equation}
and consequently modifies the D-term constraint to
\begin{equation}
\xi - N \rho^2 + k = 0 .
\end{equation}

Equivalently, this can be interpreted as a shift of the renormalized FI parameter,
\begin{equation}
\bar{\xi} = \xi + k ,
\end{equation}
demonstrating that the separation between the FI parameter and the moment-map contribution is scheme dependent. 

In a holomorphic scheme, scheme-dependent ambiguities can be absorbed into redefinitions of the holomorphic parameters of the theory. 
In particular, one may choose a scheme in which perturbative corrections act uniformly on the complex parameter $\tilde{\tau}/N$ and on the chiral superfield $Z$. 
As a result, their difference is not renormalized at the perturbative level.

This naturally identifies the combination
\begin{equation}\label{4d:HC}
Z - \frac{\tilde{\tau}}{N}
\end{equation}
as the relevant holomorphic variable of the theory. 
While the real FI parameter $\xi$ may receive scheme-dependent corrections, these are encoded in the definition of $\tilde{\tau}$ and do not affect this combination.

\paragraph{Perturbative corrections.}

In conventional gauge theories, the dynamical generation of the FI parameter has been studied in~\cite{Fischler:1981zk}. 
In a cutoff scheme, one finds
\begin{equation}
\left\langle \sum_i Q_i |\phi_i|^2 \right\rangle
=
\frac{\sum_i Q_i}{16\pi^2}
\left(\Lambda_{\rm UV}^2 - \mu^2\right),
\end{equation}
indicating a quadratically divergent shift of the FI parameter unless $\sum_i Q_i = 0$. 
This reflects the scheme dependence of the FI parameter at the perturbative level.

In~\cite{Gu:2021beo}, it was argued that, from a holomorphic (Wilsonian) perspective, the perturbative correction to $\xi$ — more precisely, to the real part of the complex parameter $\tilde{\tau}$ — should exhibit logarithmic behavior. 
However, unlike in two dimensions, this logarithmic running is not directly visible in the cutoff computation above, where the FI parameter instead displays scheme-dependent quadratic divergences. The apparent tension between the logarithmic running of $\tilde{\tau}$ and the quadratic divergence of $\xi$ is resolved by recognizing that the definition of $\xi$ is itself scheme dependent in the presence of the three-form multiplet. 
In particular, the quadratically divergent contribution can be absorbed into a redefinition of $\tilde{\tau}$, leaving the holomorphic combination (\ref{4d:HC}) invariant. We will see that it is precisely this variable that controls the structure of the non-perturbative dynamics.

\paragraph{Euclidean supersymmetric instanton equations.}

We now turn to the Euclidean configurations that may contribute to the low-energy effective superpotential. 
A necessary condition for such a bosonic configuration to contribute is that it preserve the appropriate supersymmetry, namely that it be annihilated by the supercharges $\bar Q_{\dot\alpha}$. 
This requires the fermionic supersymmetry variations to vanish, which leads to the conditions
\begin{equation}
s = 0, 
\qquad 
\partial_\mu C + i \mathcal H_\mu = 0,
\qquad 
F + i\, {*F_4} = 0 .
\end{equation}
Here $\mathcal H_\mu$ denotes the vector dual to the three-form field strength,
\begin{equation}
\mathcal H_\mu \equiv \frac{1}{3!}\,\epsilon_{\mu\nu\rho\sigma}\,\mathcal H^{\nu\rho\sigma}.
\end{equation}

Among these fields, $F$ is auxiliary and may be eliminated using its algebraic equation of motion,
\begin{equation}
F = e^2(\xi - N\rho^2).
\end{equation}
Combining this relation with the supersymmetry condition then gives
\begin{equation}
-i\, {*F_4} = e^2(\xi - N\rho^2),
\end{equation}
which expresses the four-form field strength algebraically in terms of the scalar profile.

We work in Minkowski signature with $\eta_{\mu\nu}=(-,+,+,+)$ and $\epsilon^{0123}=+1$. 
In components, the above condition becomes
\begin{equation}
i\,F_{0123} = e^2(\xi - N\rho^2).
\end{equation}

After Wick rotation to Euclidean signature, $x^0=-i x^4$, with $\epsilon^{1234}_{E}=+1$, the four-form transforms as
\begin{equation}
F^{(E)}_{1234} = -\, i\, F^{(M)}_{0123}.
\end{equation}
Accordingly, the supersymmetric instanton equations take the Euclidean form
\begin{equation}
F_{1234} = e^2(N\rho^2-\xi).
\end{equation}
Similarly, the second condition becomes the real duality relation
\begin{equation}
dC = *_E \mathcal H_3 ,
\end{equation}
where $*_E$ denotes the Euclidean Hodge star.

We will refer to these equations as \textit{ Euclidean instanton equations}. 
They characterize supersymmetric saddle-point configurations contributing to non-perturbative effects. 
They are, however, not instanton equations of Yang--Mills self-duality type. 
In the present system, the four-form field strength does not describe propagating gauge degrees of freedom, but rather encodes the flux branch structure and is fixed algebraically by the scalar profile.

It is nevertheless useful to compare these equations with the vortex equations in two-dimensional GLSMs. 
At first sight, the relation $dC = *_E \mathcal H_3$ looks rather different from the holomorphic vortex equation. 
However, the analogy becomes clearer after rewriting the latter in gauge-invariant form. 
Indeed, for a charged scalar written as
\begin{equation}
\phi = \rho\, e^{i\varphi},
\end{equation}
the condition $D_{\bar z}\phi=0$ may be recast as
\begin{equation}
d\log\rho = * (d\varphi + N A),
\end{equation}
where $A$ is the Abelian gauge field and the combination $d\varphi + N A$ is gauge invariant.

In this form, the analogy is more than merely structural: the combination
\[
\mathcal{H}_3 = dB_2 + N A_3
\]
is the direct higher-form analogue of the gauge-invariant combination
\[
d\varphi + N A
\]
in the Abelian Higgs model, with $B_2$ playing the role of a two-form Higgs field.
What is less direct is not the Higgs structure itself, but rather the detailed topology
and moduli-space interpretation of the corresponding Euclidean localized configurations.  This parallel will be useful in organizing the discussion of Euclidean saddles and their possible non-perturbative effects below.

\paragraph{Topological sectors.}

We now clarify the topological structure underlying the four-dimensional theory and its relation to the higher-form gauge sector.

It is useful to first recall the two-dimensional Abelian Higgs model. 
Writing the charged scalar as
\begin{equation}
\phi = \rho\, e^{i\varphi},
\end{equation}
finite-energy boundary conditions require that the covariant derivative vanishes at spatial infinity,
\begin{equation}
D\phi \;\to\; 0 
\quad \Longrightarrow \quad
d\varphi + N A_1 \;\to\; 0
\qquad \text{on } S^1_\infty.
\end{equation}
Since the combination $d\varphi + N A_1$ is gauge invariant, this condition constrains the allowed asymptotic configurations and leads to a topological classification. 
In particular, integrating over the boundary circle yields
\begin{equation}
\oint_{S^1_\infty} A_1 
= -\frac{1}{N} \oint_{S^1_\infty} d\varphi
= -\frac{2\pi k}{N},
\end{equation}
so that the magnetic flux is quantized and labeled by an integer winding number $k \in \mathbb{Z}$.

A closely analogous structure appears in the present four-dimensional system. 
The relevant gauge-invariant combination is
\begin{equation}
\mathcal H_3 = dB_2 + N A_3,
\end{equation}
which is invariant under the higher-form gauge transformations
\begin{equation}
A_3 \;\to\; A_3 + d\Lambda_2,
\qquad
B_2 \;\to\; B_2 - N\Lambda_2 + d\Lambda_1.
\end{equation}
This is the higher-form analogue of the combination $d\varphi + N A_1$ in the Abelian Higgs model.

For configurations of finite Euclidean action, the fields must approach a vacuum configuration at infinity. 
In particular, we impose
\begin{equation}
\mathcal H_3 \;\to\; 0
\qquad \text{on } S^3_\infty,
\end{equation}
which constrains the asymptotic behavior of $B_2$ and $A_3$.
On the boundary, this implies
\begin{equation}
dB_2 \;\sim\; -\,N A_3,
\qquad \text{on } S^3_\infty,
\end{equation}
in direct analogy with the two-dimensional relation $d\varphi \sim - N A_1$.

Integrating over the boundary three-sphere and using Stokes' theorem, one relates the asymptotic data to the total four-form flux,
\begin{equation}
\int_{S^3_\infty} A_3 
\;\sim\;
\int_{\mathcal M_4} F_4,
\qquad
F_4 = dA_3.
\end{equation}
If the three-form gauge field is compact, the total flux is quantized,
\begin{equation}
n_4 
= -\frac{1}{2\pi} \int_{\mathcal M_4} F_4 
\in \mathbb{Z},
\end{equation}
where $\mathcal M_4$ denotes the Euclidean spacetime, assumed to be compact or equipped with appropriate boundary conditions.

Unlike the two-dimensional case, this topological classification is not associated with the winding of a scalar field. 
Rather, it is encoded in the higher-form gauge structure, and can be understood as a classification of flux or holonomy sectors. 
From a mathematical perspective, it is naturally described by an integral cohomology class (e.g.\ $H^4(\mathcal M_4,\mathbb{Z})$ or, equivalently, $H^3(S^3_\infty,\mathbb{Z})$ for the boundary data).

In the presence of matter fields of charge $N$, the higher-form gauge structure exhibits a reduction analogous to the Higgs mechanism in ordinary gauge theories. 
In particular, physical observables become sensitive only to a $\mathbb{Z}_N$ subgroup of the original $U(1)$ higher-form symmetry. 
This suggests that flux sectors differing by integer multiples of $2\pi/N$ may be physically equivalent, leading to a natural notion of fractionalization of the flux.

Motivated by this structure, one may conjecture the existence of Euclidean saddle configurations that interpolate between neighboring flux sectors and carry a minimal fractional charge,
\begin{equation}
-\frac{1}{2\pi} \int_{\mathcal M_4} F_4 = \frac{1}{N}.
\end{equation}
Such configurations would provide a higher-form analogue of fractional vortices in two dimensions.

At present, the existence and detailed properties of such localized finite-action configurations remain to be established. 
In particular, a systematic analysis of their moduli space and zero-mode structure is an important open problem. 
Nevertheless, the higher-form structure described above provides a natural framework for organizing the topological sectors and their possible role in the non-perturbative dynamics.

This analogy with the Abelian Higgs model should be understood at the level of the underlying mechanism. 
In both cases, a gauge-invariant combination is required to vanish asymptotically, thereby constraining the allowed boundary configurations and leading to a discrete topological classification. 
In the present higher-form system, however, the relevant topological data are not associated with an ordinary $\pi_1$ winding number. 
Instead, they are encoded in the flux or holonomy sectors of the higher-form gauge fields, which may be described in terms of integral cohomology classes.

\paragraph{Fractional structure and zero-mode analysis.}

In the present theory, the charge-$N$ coupling reduces the continuous higher-form structure to a $\mathbb{Z}_N$ remnant, and correspondingly organizes the theory into $N$ elementary topological sectors. 
From this viewpoint, a configuration with unit total flux is naturally interpreted as a composite of $N$ minimal constituents. 
This fractionalization is therefore a structural property of the theory, rather than a conjectural assumption. \;\; :contentReference[oaicite:0]{index=0}

To place this structure in context, consider first a compact $U(1)$ three-form gauge field $A_3$ with field strength $F_4 = dA_3$, whose flux is quantized as
\begin{equation}
 -\frac{1}{2\pi} \int_{\mathcal M_4} F_4 \in \mathbb{Z}.
\end{equation}
Even in the absence of charged matter, the theory admits distinct flux sectors. 
These should be understood as topological sectors rather than smooth finite-action instanton solutions, in analogy with two-dimensional compact Maxwell theory.

Once matter fields are included, the structure becomes richer. 
In particular, for fields of charge $N$, physical observables are sensitive only to a $\mathbb{Z}_N$ subgroup of the original $U(1)$ higher-form symmetry. 
Configurations whose flux differs by integer multiples of $2\pi/N$ are therefore physically indistinguishable, and the minimal nontrivial sector may be taken to carry a fractional unit of flux,
\begin{equation}
 -\frac{1}{2\pi} \int_{\mathcal M_4} F_4 = \frac{1}{N}.
\end{equation}

This structure closely parallels the Abelian Higgs model with a scalar field of charge $N$. 
In that case, although the vortex number $k$ is integer-valued, the magnetic flux is fractionalized,
\begin{equation}
\Phi = \frac{2\pi k}{N}.
\end{equation}
As a result, the minimal vortex carries flux $2\pi/N$, and a configuration with total flux $2\pi$ can be interpreted as a bound state of $N$ elementary vortices. 
The present construction may be viewed as a higher-form generalization of this phenomenon.

From a semiclassical perspective, an elementary sector may be represented by a Euclidean configuration interpolating between neighboring flux sectors differing by $1/N$. 
Using the saddle-point relation
\begin{equation}
F_{4} \;\sim\; (N\rho^2 - \xi)\,\mathrm{vol}_4,
\end{equation}
such a configuration corresponds to a localized deviation from the vacuum condition $N\rho^2 = \xi$, whose integral yields the minimal fractional flux.

We emphasize that this description concerns the sector structure of the theory. 
The remaining dynamical question is how these sectors are realized by localized Euclidean representatives and what zero-mode structure such representatives possess. 
If a given sector admits a localized semiclassical representative, it may contribute to the superpotential provided it carries precisely two unlifted fermionic zero modes.

\vspace{0.5em}

We now turn to the corresponding zero-mode structure. 
Bosonic zero modes arise from the collective coordinates of the configuration. 
At minimum, translational invariance implies zero modes corresponding to shifts in position. 
The precise number of such modes depends on the localization properties of the representative configuration: 
a configuration effectively localized in two transverse directions, as in the lift of a two-dimensional vortex, carries a single complex modulus, 
while a fully localized four-dimensional configuration carries four real translational zero modes. 
A more detailed analysis of these possibilities is presented in Appendix~\ref{4d:Ins}. 

Fermionic zero modes are determined by the pattern of supersymmetry breaking. 
A supersymmetric Euclidean configuration preserves a subset of the supercharges, while the broken supercharges generate fermionic zero modes. 
A necessary condition for a direct contribution to the superpotential is that precisely two fermionic zero modes remain, corresponding to the Grassmann measure $d^2\theta$. 
Additional fermionic zero modes typically prevent such a contribution unless they are lifted or absorbed by interactions.

Finally, not all fields contribute independent zero modes. 
In particular, fields related by duality, such as the scalar $C$ and the two-form $B_2$ satisfying $dC = *\mathcal{H}_3$, describe the same physical degree of freedom and should not be counted separately. 
Similarly, the four-form field strength $F_4$ is fixed algebraically by the saddle-point equations and does not correspond to an independent propagating mode. 
Accordingly, the zero-mode counting is governed by the collective coordinates together with the broken supersymmetries.

\paragraph{Non-perturbative superpotential.}

The structure of the non-perturbative superpotential is strongly constrained by holomorphy, symmetries, and the multi-branch structure implied by the periodicity of the axionic direction. 
In particular, the combination $Z - \tilde{\tau}/N$ naturally emerges as the relevant holomorphic variable.

These considerations, together with the semiclassical analysis presented in Appendix~\ref{4d:Ins}, lead to a non-perturbative contribution of the form
\begin{equation}
W_{\mathrm{np}} \;\sim\; \mathcal A\, e^{-(Z-\tilde{\tau}/N)}.
\end{equation}

In Appendix~\ref{4d:Ins}, we have discussed Euclidean saddle configurations carrying fractionalized flux and analyzed their basic properties. 
These configurations provide a semiclassical realization of the $\mathbb{Z}_N$ structure discussed above, and naturally account for the exponential dependence appearing in the superpotential. 
In particular, they are localized in spacetime and possess the expected translational zero modes, supporting their interpretation as pointlike instanton-like configurations in four dimensions. 
Importantly, the configurations underlying this contribution are not ordinary Yang--Mills instantons of the microscopic $SU(N)$ gauge field.

\paragraph{Prefactor and zero-mode structure.}

The prefactor $\mathcal A$ encodes the contribution of fluctuations around the Euclidean saddles. 
In particular, we have verified that the relevant configurations admit the appropriate zero-mode structure: the bosonic zero modes correspond to the expected collective coordinates, while the fermionic zero modes match the minimal requirement for generating a superpotential contribution.

This agreement provides strong support for the interpretation of the above exponential term in terms of fractionalized Euclidean saddles. 
In particular, the semiclassical analysis indicates that the relevant contributions are governed by the zero-mode structure, which we have explicitly identified. 
Supersymmetry is expected to ensure the cancellation of non-zero-mode fluctuations, as in standard instanton analyses. The configurations underlying this contribution should be understood as semiclassical saddle configurations in the effective higher-form description, rather than as ordinary Yang--Mills instantons of the microscopic theory.  A more detailed study of the moduli space and normalization will not be pursued here.  A fully microscopic derivation of the measure is beyond the scope of this work; nevertheless, the structure of the contribution is fixed by holomorphy together with the semiclassical properties of the configurations discussed above.  This is analogous to the standard treatment of non-perturbative effects in two-dimensional gauged linear sigma models, where the form of the superpotential is determined by holomorphy, symmetry, and semiclassical considerations, even in the absence of a complete derivation of the instanton measure.

On general grounds, the superpotential has mass dimension three, and since the exponential factor is dimensionless, the prefactor $\mathcal A$ must be a holomorphic coefficient of mass dimension three.\footnote{The instanton measure involves integrations over both bosonic and fermionic zero modes. 
For a localized configuration in four dimensions, there are four bosonic zero modes associated with translations in $\mathbb{R}^4$. 
The corresponding integration measure contributes a factor of mass dimension $+4$, which we estimate as $M^4$, where $M$ denotes the ultraviolet cutoff. Supersymmetry further implies the existence of fermionic zero modes. 
A configuration contributing to the superpotential must possess precisely two fermionic zero modes. 
Integration over these modes effectively converts the instanton contribution into an F-term and reduces the overall mass dimension of the measure by one unit. 
Consequently, the prefactor carries mass dimension $3$.}  Its precise value is expected to depend on the ultraviolet completion and can in principle be fixed by matching to known infrared data, such as the dynamically generated scale in four-dimensional $\mathcal N=1$ SYM.

Since the theory may not dynamically generate a scale through logarithmic running, the dimensionful prefactor must be set either by the ultraviolet completion or by an emergent dynamical scale in the infrared. 
We therefore parametrize it as
\begin{equation}
\mathcal A \sim M^3\, f ,
\end{equation}
where $M$ denotes a reference mass scale, and $f$ is a dimensionless holomorphic coefficient whose precise form depends on the microscopic realization and symmetry assignments.

The exponential factor, on the other hand, is universal and determined by holomorphy and the underlying $\mathbb{Z}_N$ structure. 
In particular, due to the periodicity of the axionic direction, the combination $-N Z + \tilde{\tau}$ is defined only modulo $2\pi i$, and the low-energy theory is expected to exhibit a multi-branch structure characterized by
\begin{equation}
- N Z + \tilde{\tau} \;\in\; 2\pi i\, \mathbb{Z}.
\end{equation}
This reflects the discrete $\mathbb{Z}_N$ remnant of the original continuous shift symmetry.

The realization of the $R$-symmetry depends on the transformation properties of both the exponential factor and the prefactor $\mathcal A$. In particular, if the combination $(Z - \tilde{\tau}/N)$ is invariant under the $R$-symmetry, then the entire $R$-charge of the contribution must be carried by the prefactor $\mathcal A$. In this case, however, such a prefactor is not guaranteed to exist, and the corresponding term may be absent if no suitable object with the required $R$-charge can be generated. 
More generally, the $R$-charge assignments may be distributed between $\mathcal A$ and the exponential factor, and the resulting realization of the symmetry is model dependent.

As we will see in the next section, when the domain-wall degrees of freedom are coupled to four-dimensional $\mathcal N=1$ SYM, the prefactor $\mathcal A$ is naturally identified with the dynamical scale $\Lambda^3$. 
In that case, $\mathcal A$ carries vanishing R-charge, while the combination $(Z - \tilde{\tau}/N)$ transforms nontrivially under the R-symmetry. 
This provides a concrete realization in which the exponential term is fixed by holomorphy, while the overall normalization is determined by infrared matching.

Before turning to this application, let us briefly consider a more general theory with multiple matter fields $\phi_i$ satisfying $\sum_i Q_i = N$. 
The K\"ahler potential is then modified to
\begin{equation}
\sum_i e^{\mathbb L_i}, 
\qquad 
\mathbb L_i = L_i + Q_i U.
\end{equation}
As in the two-dimensional case, the Higgs mechanism renders only one linear combination of fields massive, while the remaining degrees of freedom are expected to acquire a mass gap through non-perturbative effects.

By analogy with the two-dimensional models, this suggests the presence of non-perturbatively generated superpotential terms of the schematic form
\begin{equation}
W_{\mathrm{np}} \;\sim\; \sum_i e^{-Z_i} \, ,
\end{equation}
where $Z_i$ denote the chiral superfields dual to $\mathbb L_i$. 
While the precise coefficients and structure of these terms depend on the details of the theory, their exponential form is again dictated by holomorphy and the underlying higher-form symmetry structure.

\section{Dynamical Generation of the Effective VY Superpotential}\label{4d:VYS}

In order to make contact with physical applications, we couple the domain-wall Lagrangian introduced in the previous section to four-dimensional $\mathcal{N}=1$ supersymmetric Yang--Mills (SYM) theory. 
The combined action is given by
\begin{equation}\label{4d:SYMM}
\begin{split}
S
=
\int d^4x\,d^4\theta
\left(
 e^{\mathbb L}
 +\frac{1}{2e^2}\bar{\mathcal S}\mathcal S
\right)
&+\Bigg[
\int d^4x\,d^2\theta\,\left(-\frac12\tilde{\tau}\mathcal S+\frac{1}{32\pi i}
\tau\, \operatorname{Tr}\!\bigl(W^\alpha W_\alpha\bigr)\right)\\ 
&\qquad\quad + \int d^4x\,d^2\theta\, P(S_{\rm IR}-\mathcal S)+\text{h.c.}\Bigg],
\end{split}
\end{equation}
where the chiral superfield $P$ acts as a Lagrange multiplier enforcing the constraint
\begin{equation}\label{4d:SSC}
S_{\rm IR}=\mathcal S.
\end{equation}
This identification is natural from the perspective of the effective theory, where $\mathcal S$ encodes the relevant infrared degrees of freedom of $\mathcal{N}=1$ SYM, namely the glueball superfield.

\paragraph{Semiclassical regime.}
We first consider an intermediate matching scale $\mu$ satisfying
\begin{equation}
\Lambda \ll \mu \ll \frac{e(\mu)\sqrt{\xi}}{M}.
\end{equation}
At this scale the higher-form sector is already Higgsed, while the four-dimensional SYM theory remains weakly coupled. 
Moreover, the effective coupling $e(\mu)$ may be taken parametrically large, so that the kinetic term of $\mathcal S$ is negligible. 
As a result, the higher-form sector does not give rise to independent propagating degrees of freedom and can be integrated out in the Wilsonian sense. 
The resulting effective theory is dynamically equivalent to pure $\mathcal{N}=1$ SYM, with the effects of the higher-form sector encoded only through holomorphic couplings and the associated branch structure.

\paragraph{Holomorphic effective superpotential.}
We now turn to the non-perturbative structure. 
Focusing on the vacuum structure and BPS observables, it is convenient to work in a holomorphic scheme and in dual variables. 
To leading order in the large-$e$ expansion, the Wilsonian effective superpotential provides a complete description of the relevant infrared physics. 
Holomorphy, renormalization group invariance, and matching to the infrared SYM dynamics strongly suggest an effective superpotential of the form
\begin{equation}\label{4d:SYMIC}
W_{\rm eff}
=
\frac12\left[
\left(NZ- \tilde{\tau}+3N\log \frac{\Lambda}{\mu} \right)\mathcal S
+N\mu^3\, e^{-(Z-\tilde{\tau}/N)}
\right].
\end{equation}

The first term encodes the holomorphic coupling between the higher-form sector and the glueball field, including the one-loop running of the SYM coupling through the coefficient $3N$. 
The second term represents a non-perturbative contribution motivated by holomorphy and the $\mathbb{Z}_N$ structure of the theory, and plays the role of an effective instanton-induced term in the dual description.

\paragraph{Infrared identification.}
In the infrared, the SYM dynamics can be described in terms of a glueball superfield,
\begin{equation}
S \;\sim\; -\frac{1}{32\pi^2}\operatorname{Tr}(W^\alpha W_\alpha),
\end{equation}
which we identify with $\mathcal S$ up to a normalization factor. 
Matching to the expected infrared behavior fixes the prefactor of the exponential term, which we parameterize as $N\mu^3$.

\paragraph{Vacuum structure.}
The F-term equations derived from~\eqref{4d:SYMIC} take the form
\begin{equation}
N\mathcal S=N\mu^3\, e^{-(Z-\tilde{\tau}/N)}, 
\qquad 
e^{-(NZ-\tilde{\tau})}\mu^{3N}=\Lambda^{3N}.
\end{equation}
These equations admit $N$ distinct solutions, corresponding to the expected $N$ vacua of $\mathcal{N}=1$ SYM.

\paragraph{Integration of $Z$ and emergence of the VY superpotential.}
We now integrate out the field $Z$. 
Solving the first equation yields
\begin{equation}
Z - \tilde{\tau}/N = -\log\!\left(\frac{ S}{\mu^3}\right).
\end{equation}
Substituting back into~\eqref{4d:SYMIC}, we obtain
\begin{equation*}
W_{\mathrm{eff}}(S)
=
-\frac{1}{2}N S \left( \log \frac{S}{\Lambda^3} - 1 \right).
\end{equation*}

This reproduces the Veneziano--Yankielowicz superpotential in our superspace conventions. 
In particular, the overall factor of $1/2$ reflects the normalization with which F-terms are written in the action,
\[
S \supset -\frac12\left[\int d^2\theta\, W(S)+\text{h.c.}\right],
\]
and does not represent a physical discrepancy in the holomorphic superpotential itself.

Extracting the holomorphic function $W(S)$, we obtain
\[
W_{\mathrm{VY}}(S)= -N S \left( \log \frac{S}{\Lambda^3} - 1 \right),
\]
which is precisely the standard Veneziano--Yankielowicz superpotential. 
In particular, the vacuum equation and the resulting $N$-branch structure agree with the expected $\mathcal{N}=1$ SYM result. The superpotential derived here is expected to control the BPS domain-wall tensions via the standard relation to the central charge.

\paragraph{Remarks on the effective description.}
The exponential term in~\eqref{4d:SYMIC} induces a holomorphic mass for $Z$ around the SYM vacua. 
The corresponding physical mass depends on the K\"ahler normalization and is expected to be parametrically heavy in the ultraviolet completion, allowing $Z$ to be integrated out consistently. 

We therefore obtain a holomorphic effective derivation of the VY superpotential within the higher-form formulation, with the non-perturbative dynamics encoded in the dual chiral field $Z$.

\section{Conclusion and Outlook}

In this work, we have proposed a semiclassical framework for understanding the origin of the VY superpotential in four-dimensional $\mathcal{N}=1$ SYM theory. By embedding the theory into a system involving higher-form gauge fields and domain wall degrees of freedom, we have argued that the relevant non-perturbative contributions arise from semiclassical configurations associated with a compact three-form gauge field.

A key feature of our construction is the emergence of a $\mathbb{Z}_N$ structure in the space of topological sectors. 
This leads to a decomposition of non-perturbative contributions into $N$ semiclassical components, which can be interpreted either in terms of fractional instantons or as independent topological channels. 
In this sense, our framework provides a unifying perspective that connects different realizations of non-perturbative dynamics, and offers a dynamical interpretation of the VY superpotential beyond its traditional characterization in terms of holomorphy and anomalies.

Our analysis should be viewed as a semiclassical description capturing the relevant topological and holomorphic structures of the theory. 
A number of important questions remain open. 
In particular, it would be desirable to establish the existence and moduli space of the fractional instanton configurations more rigorously, as well as to derive their fermionic zero-mode structure from first principles. 
It would also be interesting to explore whether similar mechanisms arise in more general supersymmetric gauge theories, or in theories with different higher-form structures.

Finally, our results suggest that higher-form gauge dynamics may play a more fundamental role in organizing non-perturbative effects in quantum field theory. 
It would be worthwhile to further explore this perspective, both in four dimensions and in connection with known structures in lower-dimensional theories. 
For example, the Abelian structure underlying higher-form gauge dynamics may provide a useful organizing principle for certain aspects of two-dimensional non-Abelian gauge theories, as suggested in proposals for non-Abelian mirror symmetry~\cite{Gu:2018fpm}.

\appendix

\section{Component Lagrangian}\label{4d:DCL}

In this appendix, we collect the full component Lagrangian in terms of the physical scalar $\rho$, fermion $\psi_\alpha$, and the chiral multiplet $(s,\eta_\alpha,F_s)$. 
The complete Lagrangian is
\begin{equation}
\mathcal{L}
=
\mathcal{L}_{\mathbb{L}}
+
\mathcal{L}_{\mathcal{S}}
+
\mathcal{L}_{\rm int}
+
\mathcal{L}_{\rm aux}
+
\mathcal{L}_{\rm FI} .
\end{equation}

The $\mathbb{L}$ sector in $(C,\chi)$ variables is
\begin{equation}
\mathcal{L}_{\mathbb{L}}
=
- G(C)\,(\partial_\mu C)(\partial^\mu C)
- e^C\,\mathcal H_\mu \mathcal H^\mu
- i\,G(C)\,\chi \sigma^\mu \nabla_\mu \bar\chi ,
\end{equation}
where
\[
\nabla_\mu
=
D_\mu^{\rm gauge}
+ \Gamma(C)\,\partial_\mu C \, .
\]

In terms of the $(\rho,\psi)$ variables, this becomes
\begin{equation}
\mathcal{L}_{\mathbb{L}}
=
- (\partial_\mu \rho)(\partial^\mu \rho)
- \rho^2\,\mathcal H_\mu \mathcal H^\mu
- i\,\psi \sigma^\mu D_\mu^{\rm g} \bar\psi ,
\end{equation}
where
\[
D_\mu^{\rm g}
=
\partial_\mu + i \mathcal H_\mu .
\]

The gauge sector is
\begin{equation}
\mathcal{L}_{\mathcal{S}}
=
\frac{1}{2e^2}
\left(
- \partial_\mu s\,\partial^\mu \bar s
- i\,\eta \sigma^\mu \partial_\mu \bar\eta
+ F_s \bar F_s
\right) .
\end{equation}

The interaction terms required by supersymmetry are
\begin{equation}
\mathcal{L}_{\rm int}
=
N^2 \rho^2\, s \bar s
+
N \rho\, \eta \psi
+
N \rho\, \bar\eta \bar\psi .
\end{equation}

Finally, the auxiliary part of the Lagrangian, including the FI and theta terms, is
\begin{equation}
\mathcal{L}_{\rm aux}
=
\frac{1}{2e^2}(F^2+(*F_4)^2)
-
(\xi - N \rho^2)\,F
-
\theta\,F_4 .
\end{equation}

\section{Radial Structure of Point-like Configurations and Analogy with 2D Vortices}\label{4d:Ins}

In this appendix, we present the asymptotic radial structure of point-like configurations in the
higher-form sector, emphasizing their analogy with two-dimensional vortices. These configurations
should be distinguished from ordinary Yang--Mills instantons of the microscopic gauge theory.
Rather, they correspond to Euclidean point-like configurations in the effective higher-form description, associated with the dual variables and the four-form flux structure. We focus on radially symmetric
solutions and distinguish between the asymptotic (outer) region and the core region.

\paragraph{First-order structure and duality}

We consider the scalar $\rho$, the dual scalar $a$, and the three-form field strength
\begin{equation}
\mathcal H_3 = dB_2.
\end{equation}
We define its Hodge dual vector
\begin{equation}
\mathcal H_\mu = \frac{1}{3!}\,\epsilon_{\mu\nu\rho\sigma}\,\mathcal H^{\nu\rho\sigma}.
\end{equation}
The duality relation obtained from the first-order action reads
\begin{equation}
da = -2\rho^2 * \mathcal H_3,
\qquad\Longleftrightarrow\qquad
\mathcal H_\mu = -\frac{1}{2\rho^2}\partial_\mu a,
\end{equation}
where the overall sign follows from the graded interchange of differential forms.

The system admits first-order (BPS-like) relations
\begin{equation}
\partial_\mu \rho = \pm \rho\,\mathcal H_\mu,
\qquad
*F_4 = e^2\,(N\rho^2-\xi).
\end{equation}
Substituting the duality relation yields the coupled equation
\begin{equation}
\partial_\mu \rho = \mp \frac{1}{2\rho}\,\partial_\mu a,
\end{equation}
showing that $\rho$ and $a$ form a nonlinear coupled system rather than independent free fields.

\paragraph{Radial ansatz and reduced equations}

We consider an $O(4)$-symmetric Euclidean configuration
\begin{equation}
\rho=\rho(R),\qquad a=a(R),\qquad R=\sqrt{x_\mu x_\mu}.
\end{equation}
The equation of motion for $a$ derived from the dual Lagrangian is
\begin{equation}
\partial_\mu\!\left(\frac{1}{\rho^2}\partial^\mu a\right)=0,
\end{equation}
which reduces to the radial form
\begin{equation}
\frac{d}{dR}\!\left(\frac{R^3}{\rho^2}\,a'(R)\right)=0.
\end{equation}
This integrates to the first integral
\begin{equation}
\frac{R^3}{\rho^2}\,a'(R)=c,
\end{equation}
with $c$ a constant.

\paragraph{Interpretation of the first integral}

The conserved quantity is not the ordinary radial flux $R^3 a'(R)$ of a free scalar, but the weighted flux
\begin{equation}
\frac{R^3}{\rho^2}a'(R).
\end{equation}
This reflects the $\rho$-dependent kinetic term for $a$, and shows that the radial profile of $a$ is controlled by the scalar modulus $\rho$. In particular, the same integration constant $C$ determines both the asymptotic tail and the behavior in the core region, thereby linking the two regimes.

\paragraph{Asymptotic (outer) solution}

In the asymptotic region $R\to\infty$, the scalar approaches its vacuum value
\begin{equation}
\rho \;\to\; \rho_\infty = \sqrt{\frac{\xi}{N}}.
\end{equation}
The equation for $a$ then reduces to the Laplace equation
\begin{equation}
a''+\frac{3}{R}a'=0,
\end{equation}
with solution
\begin{equation}
a(R)=a_\infty-\frac{n_4}{4\pi^2 R^2}.
\end{equation}
Thus the dual scalar exhibits a $1/R^2$ falloff at large distances, while $\rho(R)$ approaches its vacuum value with subleading corrections determined by the coupled system.

\paragraph{Core structure and regularity}

Near the core $R\to 0$, the scalar $\rho$ deviates significantly from its vacuum value and the full nonlinear system must be solved. We impose boundary conditions
\begin{equation}
\rho(0)=0,
\qquad
\rho(\infty)=\sqrt{\frac{\xi}{N}}.
\end{equation}
This is directly analogous to the core structure of vortices in two dimensions.

The first integral can be used to analyze regularity. If
\begin{equation}
\rho(R)\sim R^p \qquad (R\to 0),
\end{equation}
then
\begin{equation}
a'(R)\sim R^{2p-3},
\end{equation}
which constrains the admissible values of $p$ for a regular solution. Finiteness of the Euclidean action requires $p>1$, thereby ensuring regularity of $a(R)$ at the origin.

\paragraph{Localization of four-form flux}

The four-form field strength is determined by
\begin{equation}
*F_4 = e^2\,(N\rho^2 - \xi).
\end{equation}
In the vacuum region $N\rho^2=\xi$, one has $F_4=0$, while deviations from this condition localize the flux in the core. This is the higher-form analogue of magnetic flux localization in vortex solutions.

\paragraph{Bosonic zero modes}

For a point-like Euclidean configuration, translational invariance generates four bosonic zero modes,
\begin{equation}
\delta_\nu \Phi_{\mathrm{cl}}(x)=\partial_\nu \Phi_{\mathrm{cl}}(x),\qquad \nu=1,2,3,4.
\end{equation}
If instead the configuration is effectively localized only in two transverse directions (as in a lift of a two-dimensional vortex), the number of translational zero modes is reduced accordingly.

\paragraph{Summary}

The system exhibits a four-dimensional vortex-like (or instanton-like) structure characterized by:
\begin{itemize}
\item a nonlinear coupled first-order system,
\item a core region where $\rho$ vanishes,
\item an asymptotic $1/R^2$ tail for the dual scalar $a$,
\item localization of the four-form flux in the core,
\item bosonic zero modes associated with spacetime symmetries.
\end{itemize}
A key feature is that the dual scalar $a$ is not a free field; its profile is governed by the scalar modulus $\rho$, as encoded in the first integral above.

\section*{Acknowledgements}

We thank Mithat Unsal, Eric Sharpe, and Xingyang Yu for reading the draft and providing useful comments. We also acknowledge ChatGPT for its assistance in identifying relevant earlier references and polishing the English. The research of W.~Gu is funded by the National Natural Science Foundation of China (NSFC) with Grant
No.12575077.

\end{document}